\documentstyle[12pt]{article}

\begin{document}
\baselineskip 5mm
\renewcommand{\theequation}{\thesection.\arabic{equation}}

\newcommand{\ov}{\overline}
\newcommand{\r}{$^{[??]}$}

\newcommand{\apsect}[1]{\setcounter{equation}{0}
                      \setcounter{subsection}{0}
                      \addtocounter{section}{1}
                      \section*{Appendix\, #1}}

\title{
The Four-fermi Coupling of  \\ the Supersymmetric  \\
 Non-linear $\sigma$-model on $G/S\otimes \{U(1)\}^k$ }

\author{Shogo Aoyama\thanks{e-mail: spsaoya@ipc.shizuoka.ac.jp} \\
       Department of Physics \\
              Shizuoka University \\
                Ohya 836, Shizuoka  \\
                 Japan}
                 
\maketitle
                 
\begin{abstract}

The reducible K\"ahler coset space $G/S\otimes \{U(1)\}^k$ is discussed in a geometrical 
approach.  We derive the formula which expresses the Riemann curvature of the reducible 
K\"ahler coset space in terms of its Killing vectors.
The formula manifests the group structure of $G$. On the basis of this formula
 we establish an algebraic method to evaluate the four-fermi coupling of the supersymmetric 
non-linear $\sigma$-model on $G/S\otimes \{U(1)\}^k$ at the low-energy limit. As an 
application we consider the supersymmetric non-linear $\sigma$-model on $E_7/SU(5)\otimes 
\{U(1)\}^3$  which contains the three families of ${\bf 10} + {\bf 5^*} + {\bf 1}$ of 
$SU(5)$ as the pseudo NG fermions. The four-fermi coupling constants among diffferent 
families of the fermions are explicitly given at the low-energy limit.

\end{abstract}

\vspace{5cm}

To appear in Nuclear Physics B

\newpage

\section{Introduction}
\setcounter{equation}{0}

The non-linear $\sigma$-model on a coset space $G/H$ is a low-energy effective theory of 
the Nambu-Goldstone(NG) bosons generated when a large group symmetry $G$ at high energy 
spontaneously breaks into a small one at low energy\cite{1}. If the supersymmetry exists at 
that high energy and if it survives in the spontaneous breaking, the NG bosons are 
accompanied by superpartners, called the pseudo NG fermions. They are described by the 
supersymmetric non-linear $\sigma$-model on a coset space $G/H$ which is 
k\"ahlerian\cite{2}. In the beginning of the 80's Buchm\"uller, Love, Peccei and Yanagida 
proposed to identify the massless pseudo NG fermions with light quarks and leptons and used 
the supersymmetric non-linear $\sigma$-model as a low-energy effective theory for the grand 
unification\cite{3}\cite{4}. Recently in ref. \cite{5} this idea was  revived  to explain 
the neutrino mass observed in the SuperKamiokande experiment. They proposed the 
supersymmetric non-linear $\sigma$-model on the K\"ahler coset space $E_7/SU(5)\otimes 
\{U(1)\}^3$ as a thoery 
which naturally accomodates the three families of right-handed neutrinos. Namely the model 
contains three families of ${\bf 10} + {\bf 5^*} + {\bf 1}$ and ${\bf 5}$ of $SU(5)$ as the 
NG supermultiplet $(\phi^\alpha, \psi^\alpha )$. Interactions among the pseudo NG fermions 
take place through the four-fermi coupling 
$$
R_{\alpha\ov\sigma\beta\ov\delta}(\textstyle{\phi \over f},\textstyle{\bar \phi \over f})
  (\bar \psi^{\ov\sigma}\psi^\alpha)(\bar\psi^{\ov\delta}\psi^\beta)
$$
in which $R_{\alpha\ov\sigma\beta\ov\delta}$ is the Riemann curvature of the coset space 
and $f$ is a constant giving a mass scale. It is phenomelogically the most interesting part 
of the model. The aim of this paper is to establish a practical method to calculate the 
Riemann curvature of $E_7/SU(5)\otimes \{U(1)\}^3$.

$E_7/SU(5)\otimes \{U(1)\}^3$ is a fairly complicated coset space. The complication comes 
in twofold. Firstly the coset space is reducible\cite{6}. 
One can  calculate   the holomorphic Killing vectors and the K\"ahler potential of the 
reducible coset space. But they take more cumbersome forms than those of the irreducible 
coset space. Secondly the homogeneous group includes plural $U(1)$s as direct products. In 
such a case the complex structure of the K\"ahler coset space is not unique  and the metric 
depends on as many  free parameters as $U(1)$s. These subjects on the reducible K\"ahler 
coset space were extensively studied by the Kyoto group in ref. \cite{7}\cite{8}. The 
general method to construct the  K\"ahler potential  of the reducible coset space was 
given. The Riemann curvature may be calculated by differentiating that K\"ahler potential 
by coordinates, in principle. But it is too involved.

In this paper we employ an alternative formalism to do this more directly, which was 
proposed in ref. \cite{9}. It is based on the Killing potentials instead of the  K\"ahler 
potential, which are also characteristic for the K\"ahler coset space $G/H$ \cite{10}. In 
ref.\cite{9} the formalism was developed for the irreducible case, and the Riemann 
curvature  was given in terms of the holomorphic Killing vectors with no derivative by 
coordinates. (See eq. (4.14).)  Once given a concrete form of the holomorphic Killing 
vectors, one can directly calculate the Riemann curvature by that formula.  In this paper 
we extend this formalism to study the reducible K\"ahler coset space along the same line. 
We will be  particularly interested in the Riemann curvature at the low-energy limit $f 
\rightarrow \infty$, i.e., $R_{\alpha\ov\sigma\beta\ov\delta}(0,0)$. It gives the 
four-fermi coupling constants  at low energy, which depend on as many free parameters as 
$U(1)$s of the homogeneous subgroup. Knowing the dependence explicitly is very interesting 
from the phenomelogical point of view.

In Section 2 we summarize the geometry of the K\"ahler coset space $G/H$. In Section 3 we 
briefly explain the generalized CCWZ formalism\cite{7}\cite{8}. It enables us to construct 
the holomorphic Killing vectors, the metric and the K\"ahler potential for the reducible 
case. In Section 4 the alternative formalism based on the Killing potentials is presented 
which allows us to calculate the metric and the Riemann curvature more directly. We first 
of all review the formalism which was de\-veloped for the irreducible case\cite{9}. Then it 
will be extended to the reducible case. We  derive the general formula which expresses the 
Riemann curvature of the reducible K\"ahler coset space in terms of the holomorphic Killing 
vectors.(See eq. (4.32).)  In Section 5   the Riemann curvatures 
 of  $SU(3)/\{U(1)\}^2$ and $E_7/SU(5)\otimes \{U(1)\}^3$ are evaluated  to the leading 
order of ${1\over f}$ by this general formula.

\section{The geometry of the K\"ahler manifold }
\setcounter{equation}{0}

In this section we briefly review on the K\"ahler manifold, giving our notation.  Consider 
a real $2N$-dimensional Riemann manifold $\cal M$ with local 
coordinates
$\phi^a =(\phi^1,\phi^2,\cdots\cdots,\phi^{2N})$. The line element of the manifold is given 
by 
\begin{eqnarray}
d s^2 =  g_{ab} d \phi^a d \phi^b.
\end{eqnarray}
$\cal M$ is a K\"ahler manifold if it is endowed with 
a complex structure which is covariantly constant:
\begin{eqnarray}
 J^a_{\ b ;c} = 0,  
\end{eqnarray}
and satisfies $ J^a_{\ b} J^b_{\ c} = -\delta^a_c$.
We assume the metric $g_{ab}$ to be of type $(1,1)$, i.e.,
\begin{eqnarray}
g_{ab} = g_{cd} J^c_{\ a} J^d_{\ b}.
\end{eqnarray}
(A tensor of more general type $(r,s)$ will be discussed soon later.) 
The symplectic structure $J_{ab}$ is given by 
$
J_{ab} = g_{ac} J^c_{\ b}.
$
By (2.2) it is closed:
\begin{eqnarray}
J_{ab,c} + 
J_{bc,a}+
 J_{ca,b} = 0.   
\end{eqnarray}
When the K\"ahler manifold is a coset space $G/H$, there is a set of Killing vectors 
\begin{eqnarray}
R^{Aa} = (R^{1a},R^{2a},\cdots \cdots, R^{Da}),
\end{eqnarray}
with $D = {\rm dim} \ G$ , which represents the isometry $G$. They satisfy
\begin{eqnarray} 
{\cal L_{R^A}} R^{Ba} & = & R^{Ab} R^{Ba}_{\ \  ,b} - R^{B\ b} R^{Aa}_{\ \ ,b}
  \ \ = \ \ f^{ABC}R^{Ca}                  \\          
& & \hspace{2cm}({\rm Isometry}),    \nonumber \\              
{\cal L_{R^A}} g_{ab} & = &
 R^{Ac} g_{ab,c} + 
 R^{Ac}_{\ \ ,a}g_{cb} +
 R^{Ac}_{\ \ ,b}g_{ca}
\ \ = \ \ 0         \\
& & \hspace{2cm}
  ({\rm Killing \ condition}), \nonumber \\   
{\cal L_{R^A}} J^a_{\ b} & = &
 R^{Ac}J^a_{\ b,c} -
 R^{Aa}_{\ \ ,c} J^c_{\ b} + 
 R^{Ac}_{\ \ ,b} J^a_{\ c} \ \ =\ \  0,     
\end{eqnarray}
in which ${\cal L_{R^A}}$ is the Lie-derivative with respect to $R^A$, and $f^{ABC}$ are 
structure constants of the isometry group $G$.

 Any vector $v^a$ can be projected onto the $(1,0)$ and $(0,1)$ types by
\begin{eqnarray}
 {1 \over 2}(1 - iJ)^a_{\ b}v^b, \quad \quad \quad
\quad\quad
 {1 \over 2}(1 + iJ)^a_{\ b}v^b.    
\end{eqnarray} 
A tensor  of the $(r,s)$ type is obtained as a multi-product of these projected vectors.
We may locally
set  the complex structure to be
\begin{eqnarray}
J^a_{\ b} = 
\left(
\begin{array}{cc}
 -i\delta^\alpha_{\ \beta}  &  0 \\
           &    \\
         0      &  i\delta^{\overline \alpha}_{\ \overline \beta} \\
\end{array}\right),   
\end{eqnarray}
with $\alpha, \overline \alpha = 1,2,\cdots,N.$
Then the respective vectors in (2.9) may be written as $N$-dimensional complex vectors 
$v^\alpha$ and $v^{\overline \alpha}$. 
The line element (2.1) is written as
$$
d s^2 =   g_{ \alpha \overline \beta} 
d \phi^{\alpha} d \phi^{\overline \beta},
$$
by (2.3). The closure of the symplectic structure given by (2.4) reads 
\begin{eqnarray}
 g_{\alpha \bar \beta,\gamma} = g_{\gamma\bar \beta, \alpha}, \quad\quad
 g_{\alpha\bar\beta,\bar\gamma} = g_{\alpha\bar\gamma,\bar\beta}.
\end{eqnarray}
Then it follows that there exists a real scalar $K(\phi,\bar\phi)$, called K\"ahler 
potential such that 
\begin{eqnarray}
g_{\alpha\bar\beta} = K_{,\alpha\bar\beta}.
\end{eqnarray}
Furthermore (2.8) and (2.10) imply that the Killing vectors are holomorphic:
\begin{eqnarray}
R^{A \ov \beta}_{\ \ ,\alpha} =  0,   \quad\quad\quad
 R^{A\beta}_{\ \ ,\ov \alpha} = 0.
\end{eqnarray}
Then (2.6) and (2.7) reduce respectively to
\begin{eqnarray}
{\cal L_{R^A}} R^{B\alpha} &=&  R^{A\beta} R^{B\alpha}_{\ \  ,\beta} - R^{B\ \beta} 
R^{A\alpha}_{\ \ ,\beta}
 \ \ = \ \  f^{ABC}R^{C\alpha},   \\
  & & {\rm c.c.},  \nonumber             
\end{eqnarray}
and
\begin{eqnarray}
{\cal L_{R^A}} g_{\alpha\ov\beta} &=& 
R^A_{\ \alpha,\ov\beta} + 
R^A_{\ \ov\beta,\alpha} \ \ = \ \ 0,
\end{eqnarray}
with $R^A_{\ \alpha} = g_{\alpha\ov\beta}R^{A\ov\beta}$ and
    $R^A_{\ \ov\alpha} = g_{\beta\ov\alpha}R^{A\beta}$. 
From (2.15) we may find real scalars $M^A(\phi,\bar\phi)$, called Killing 
potentials\cite{10}, such that\rm  
\begin{eqnarray}
R^A_{\ \alpha} = \ i M^A_{\ ,\alpha} , \quad \quad
R^A_{\ \ov \alpha} = - i M^A_{\ \ov \alpha}.   
\end{eqnarray}
As shown in ref.\cite{10},
they   transform as the adjoint representation of the group $G$ by the Lie-variation
\begin{eqnarray}
{\cal L_{R^A}} M^B = R^{A\alpha}M^B_{\ ,\alpha} +
                      R^{A\ov \alpha} M^B_{\ ,\ov\alpha}
                   = f^{ABC}M^C.    
\end{eqnarray}
A manipulation of (2.17) with (2.16) leads us to write the Killing potentials in terms of 
the Killing vectors\cite{9}\cite{11} : 
\begin{eqnarray}
M^A = -{i\over {\cal N}_{adj}}  f^{ABC}R^{B\alpha}
R^{C\overline \beta} g_{\alpha \overline \beta}.   \nonumber
\end{eqnarray}
Here we have used the normalization 
\begin{eqnarray}
f^{ABC}f^{ABD} = 2{\cal N}_{adj} \delta^{CD}.
\end{eqnarray}
These Killing potentials characterize the K\"ahler manifold no less than the K\"ahler 
potential, if it is a coset space $G/H$.

\section{The CCWZ formalism}
\setcounter{equation}{0}

In this section we will explain how to construct the holomorphic Killing vectors 
$R^{A\alpha}$, the metric $g_{\alpha \ov \beta}$ and the  K\"ahler potential $K$, which 
essentially characterize the K\"ahler coset space $G/H$. When the K\"ahler coset space 
$G/H$ is irreducible, they can be constructed case by case in heuristic 
ways\cite{4}\cite{12}. But for the reducible case we need a systematic method. It was 
 given by generalizing the CCWZ formalism \cite{1} by the Kyoto group\cite{7}\cite{8}. We 
briefly sketch this generalized CCWZ formalism.
 
\subsection{The holomorphic Killing vectors} 

We assume the isometry group $G$ is compact and semi-simple. If a coset space $G/H$ is 
k\"ahlerian , the unbroken subgroup $H$ contains $U(1)$ groups as $H = S\otimes \{U(1)\}^k, 
k = 1,2,\cdots,n$, according to the Borel theorem\cite{13}. The generators $T^A$ of $G$ are 
decomposed as
\begin{eqnarray}
\{ T^A\} = \{ X^a, S^I, Q^\mu \}, \quad\quad 
     a & = & 1,2,\cdots,2N(= {\rm dim}\ G - {\rm dim}\ H),\nonumber\\
     I & = & 1,2,\cdots,{\rm dim}\ S (={\rm dim}\ H - k), \nonumber\\
     \mu & = & 1,2,\cdots,k,
\end{eqnarray}
in which $S^I$ and $Q^\mu$ are generators of $S$ and $U(1)$s respectively, while $X^a$ 
broken generators. Let us define a central charge as
\begin{eqnarray}
Y = \sum_{\mu=1}^k v^\mu Q^\mu \equiv v\cdot Q,
\end{eqnarray}
by choosing real coefficients $v^\mu$ such that all the broken generators $X^a$ have 
non-vanishing $Y$-charges. Then the broken generators $X^a$ can be splitted into two parts: 
the generators $X^{\ov i}$ with positive $Y$-charge and their hermitian conjugates $X^i$ 
with negtaive charge, $i,\ov i = 1,2,\cdots,N$. (3.1) is further decomposed as 
\begin{eqnarray}
\{ T^A\} = \{ X^{\ov i},X^i, S^I, Q^\mu \}.
\end{eqnarray}
The splitting of the broken generators determines the complex structure $J^a_{\ b}$ of the  
K\"ahler coset space $G/H$. But the splitting is not unique depending on the definition of 
the central charge (3.2). It implies arbitrariness of the complex structure of the coset 
space.

For the decompostion (3.3) the standard application of the CCWZ formalism does not give the 
holomorphic Killing vectors $R^{A\alpha}$ satisfying the Lie-algebra (2.14). Hence we 
extend the isometry group $G$ to the complex one $G^c$ and consider a coset space $G^c/\hat 
H$ with the complex subgroup $\hat H$ generated by $X^i, S^I, Q^\mu$\cite{14}. As 
explicitly given later, there is an isomorphism between this complex coset space $G^c/\hat 
H$ and $G/H$:
\begin{eqnarray} 
G/H \cong  G^c/\hat H.
\end{eqnarray}
The holomorphic Killing vectors are obtained by applying the CCWZ formalism to the complex 
coset space  $G^c/\hat H$. The coset space  $G^c/\hat H$ is parametrized by complex 
coordinates $\phi^\alpha, \alpha = 1,2,\cdots,N$ corresponding to the broken generators 
$X^i$. Consider a holomorphic quantity 
\begin{eqnarray}
\xi (\phi) = e^{\phi\cdot \bar X} \in G^c/\hat H
\end{eqnarray}
with \footnote{(3.5) should have been written as $\xi (\phi) = \exp ({1\over f}\phi\cdot 
\bar X)$ with the mass scale parameter $f$. But it is hereinafter set to be one to avoid 
unnecessary complication. } 
$$
\phi\cdot X^{\ov i} = \phi^1 X^{\ov 1} + \phi^2 X^{\ov 2}+ \cdots
 \phi^N X^{\ov N}.
$$
 For an element $g$ of the isometry group $G$, i.e.,
 $g = e^{i\epsilon^A T^A} \in G$ with real parameters $\epsilon^A$, there exists a 
compensator $\hat h (\phi,\bar \phi,g) \in \hat H$ such that 
\begin{eqnarray}
g\xi (\phi) = \xi (\phi') \hat h (\phi,\bar\phi, g).  
\end{eqnarray}
This defines a holomorphic transformation of the coordinates $\phi^\alpha$ which realizes 
the isometry group non-linearly. When the real parameters $\epsilon^A$ are infinitesimal, 
(3.6) yields the holomorphic Killing vectors $R^{A\alpha}(\phi)$  as
\begin{eqnarray}
\delta \phi = \phi'^\alpha (\phi)  - \phi^\alpha = \epsilon^A 
R^{A\alpha}(\phi),\end{eqnarray}
which satisfy the Lie-algebra (2.14).

\subsection{The metric}

Any two points on the coset space $G/H$ can be related by the isometry transformation 
(3.6). Therefore the line element (2.1) has the same length at any point of the coset space
\begin{eqnarray}
g_{ab}(\phi',\bar\phi') d\phi'^a d\phi'^b = g_{ab}(\phi,\bar\phi) d\phi^a d\phi^b. 
\end{eqnarray}
On the other hand the line element is invariant under general coordinate transformations:
\begin{eqnarray}
 g_{ab}(\phi,\bar\phi) d\phi^a d\phi^b = g'_{ab}(\phi',\bar\phi') d\phi'^a d\phi'^b.
\end{eqnarray}
(3.8) and(3.9) require that 
\begin{eqnarray}
 g_{ab}(\phi',\bar\phi') = g'_{ab}(\phi',\bar\phi')
\end{eqnarray}
which gives the Killing condition (2.7) in the infinitesimal form.

To construct the metric $g_{ab}$ which satisfy the condition (3.10) we have recourse to the 
CCWZ formalism. Consider a quantity
\begin{eqnarray}
 U(\phi,\bar \phi) \in G/H,
\end{eqnarray}
with $U^\dagger U = UU^\dagger = 1$.
But the standard parametrization of $U$, i.e, $U(\phi,\bar \phi) = e^{\phi\cdot \bar X + 
\bar \phi\cdot X}$ does not give the metric of the type (1,1). Therefore we employ the 
non-standard one, namely 
\begin{eqnarray}
 U(\phi,\bar \phi) = \xi (\phi)\zeta (\phi,\bar \phi),
\end{eqnarray}
in which $\xi (\phi)$ is the element (3.5), while $\zeta (\phi,\bar \phi)$ an element of 
$\hat H$. We para\-metrize the latter as 
\begin{eqnarray}
 \zeta (\phi,\bar \phi)= e^{a(\phi,\bar \phi)\cdot X} e^{b(\phi,\bar \phi)\cdot S} 
e^{c(\phi,\bar \phi)\cdot Q},
\end{eqnarray}
with
\begin{eqnarray}
 a\cdot X = \sum_{i=1}^N a^iX^i, \quad \quad \quad
 b\cdot S = \sum_{I=1}^{{\rm dim}H - k} b^IS^I. \nonumber
\end{eqnarray}
Here the function $b(\phi,\bar \phi)$ and $c(\phi,\bar \phi)$ are chosen to be real since 
their purely imaginary parts can be absorbed into an element of $H$. Then the 
parametrization (3.13) is determined by the unitary condition $U^\dagger U = 1$ which reads 
\begin{eqnarray}
 \xi^\dagger (\bar \phi) \xi (\phi) = e^{-\bar a (\phi,\bar \phi)\cdot \bar X}
                                      e^{-2b (\phi,\bar \phi)\cdot  S}
                                      e^{-2 c (\phi,\bar \phi)\cdot Y}
                                      e^{-a (\phi,\bar \phi)\cdot X}.
   \nonumber
\end{eqnarray}
(3.12) is an concrete expression of the isomorphism (3.4) relating the respective elements 
(3.5) and (3.11) of the coset spaces $G/H$ and $G^c/\hat H$.

The fundamental object to construct the metric $g_{\alpha\ov \beta}$ is the Cartan-Maurer 
$1$-form
\begin{eqnarray}
 \omega &=& U^{-1}d U   \nonumber \\
        &=& e^i X^{\ov i} + e^{\ov i}X^i + \omega^IS^I + \omega^\mu Q^\mu, 
\end{eqnarray}
with the $1$-forms $e^i (\phi,\bar \phi), e^{\ov i}(\phi,\bar \phi),\omega^I(\phi,\bar 
\phi)$ and $\omega^\mu (\phi,\bar \phi)$ as coefficients of the expansion. In particular  
$e^i (\phi,\bar \phi)$ takes the form 
\begin{eqnarray}
 e^i = e^i_\alpha d \phi^\alpha, \nonumber
\end{eqnarray}
with no $d \phi^{\ov\alpha}$,
as can be seen from the parametrization (3.12). $e^{\ov i} (\phi,\bar \phi)$
 is its complex conjugate. The components $e^i_\alpha$ and $e^{\ov i}_{\ov \alpha}$ are 
vielbeins of the local frame of the coset space. From this it follows that 
\begin{eqnarray}
 g_{\alpha\beta}& = & g_{\bar \alpha \bar \beta} \ \ = \ \ 0,  \nonumber\\
 g_{\alpha\bar\beta}&= & g_{\bar\beta\alpha} \ =\  \sum_{i=1}^N y_i(v)e^i_\alpha 
 e^{\ov i}_{\ov \beta},
\end{eqnarray}
where $y_i(v)$ is the positive $Y$-charge (3.2) of the broken generator $X^{\ov 
i}$\cite{7}:
\begin{eqnarray}
 [Y,X^i] &=& -y_i(v)X^i,  \quad\quad\quad
 [Y,X^{\ov i}] \ \ = \ \  y_i(v)X^{\ov i}. 
\end{eqnarray}
By the transformation (3.6) $U$ transforms as 
\begin{eqnarray}
 gU(\phi,\bar \phi) = U (\phi',\bar \phi')h(\phi,\bar \phi,g),
\end{eqnarray}
with a compensator $h \in H$\cite{8}. Then $e^i$ transform homogeneously as 
\begin{eqnarray}
 e^i (\phi',\bar \phi') = \rho^{ij}(h(\phi,\bar \phi,g),g)e^j(\phi,\bar \phi),
\end{eqnarray}
in which $\rho^{ij}(h,g)$ is the $N$-dimensional representation of the subgroup $H$. 
Consequently the metric (3.15) satifies the transformation property (3.8) under (3.6) or 
equivalently (3.17). Furthermore (3.16) guarantees the closure property of the metric 
(2.11)\cite{7}. If the K\"ahler coset space $G/H$ is reducible, the broken generators $X^i$ 
are decomposed into irreducible sets under the subgroup $H$, each of which may have a 
different $Y$-charge due to the Schur's Lemma. 

It can be also shown\cite{7}\cite{8} that one can write the metric (3.15) as (2.12) with 
the K\"ahler potential
\begin{eqnarray}
 K(\phi,\bar \phi) = \sum_{\mu =1}^k v^\mu c^\mu (\phi,\bar \phi),
\end{eqnarray}
where $c^\mu$ are the functions appearing in the parametrization (3.13) and $v^\mu$ are the 
coefficients defining the $Y$-charge (3.2).

\section{The Riemann curvature}
\setcounter{equation}{0}

The Riemann curvature of the  K\"ahler manifold is given by 
\begin{eqnarray}
 R_{\alpha\ov \sigma \beta \ov \delta}& =& g_{\eta\ov \sigma}\Gamma^\eta_{\alpha\beta,\ov 
\delta}  \\
 &=& g_{\alpha\ov\sigma,\beta\ov\delta} - g^{\kappa\ov\lambda}
  g_{\alpha\ov\lambda,\beta}g_{\kappa\ov\sigma,\ov\delta}.    \nonumber
\end{eqnarray}
To obtain it explicitly we have to compute the metric $g_{\alpha\ov \beta}$ in the first 
place. It may be done  with (3.15) by calculating the vielbeins $e^i_\alpha$ or with (2.12) 
by calculating the K\"ahler potential (3.19). Either calculation  is already complicated. 
It is further complicated  to take the derivative $g_{\alpha\ov\sigma,\beta\ov\delta}$ to 
obtain the Riemann curvature. Hence in this section we will study a method which enables us 
to calculate the Riemann curvature in a more direct way.

\subsection{The irreducible case\cite{9}}

When the K\"ahler manifold $G/H$ is irreducible, all the broken generators $X^{\ov i}$ have 
the same $Y$-charge $y(v) (>0)$. Then the metric (3.15) becomes simple:
\begin{eqnarray}
 g_{\alpha\ov \beta} = y(v) \sum_{i=1}^N e^i_\alpha 
  e^{\ov i}_{\ov \beta}, 
\end{eqnarray}
the value of which at the origin of the manifold is
\begin{eqnarray}
 g_{\alpha \ov \beta}\mathop{ \vert}_{\phi = \bar \phi = 0}
    = y(v) \delta_{\alpha \ov \beta}.
\end{eqnarray}
It was the Killing condition (2.14) that allows us to write the metric in the form of 
(4.2). The Killing condition can be satisfied also by giving the metric in terms of the 
Killing vectors (2.5): $g^{\alpha\ov \beta} \propto R^{A\alpha}R^{A\ov \beta}$. Fixing the 
free parameter by the initial condition (4.3) we then have 
\begin{eqnarray}
 g^{\alpha\ov \beta} = {1\over y(v)} R^{A\alpha}R^{A\ov \beta},
\end{eqnarray}
which should be equivalent to the metric given by (4.2). Here we have used
\begin{eqnarray}
 R^{A\alpha} \mathop{\vert}_{\phi = 0} = i\delta^{A\alpha}, \quad\quad\quad
 R^{A\ov\alpha}\mathop{ \vert}_{\ov\phi = 0} = -i\delta^{A\ov\alpha}, \quad\quad\quad
\end{eqnarray}
which are obvious by the construction in Subsection {\bf 3.1}.  For other components of the 
metric we have 
\begin{eqnarray}
 g^{\alpha \beta} = R^{A\alpha}R^{A\beta}=0, \quad\quad\quad g^{\ov\alpha\ov \beta} = 
R^{A\ov\alpha}R^{A\ov\beta} = 0.
\end{eqnarray}
With (4.4) the Affine connection becomes 
\begin{eqnarray}
 \Gamma_{\alpha\beta}^{\ \ \eta} &=& g^{\eta\ov \sigma}g_{\alpha\ov\sigma,\beta}
    \ \ = \ \ - g^{\eta\ov \sigma}_{\ \  ,\beta}g_{\alpha\ov\sigma} \nonumber\\
    &=& -{1\over y(v)}R^A_{\ \beta} R^{A\eta}_{\ \ , \alpha}
    \ \ = \ \ -{1\over y(v)}R^A_{\ \alpha} R^{A\eta}_{\ \ \ , \beta} .
\end{eqnarray}
by using the property (2.11).
Putting this into (4.1) we have 
\begin{eqnarray}
 R_{\alpha\ov\sigma\beta\ov\delta} 
   &=& g_{\eta\ov\sigma}
     (-{1\over y(v)}R^A_{\ \beta,\ov\delta}R^{A\eta}_{\ \ \ ,\alpha}  ) \nonumber\\
   &=& g_{\eta\ov\sigma}
     (-{1\over y(v)}R^A_{\ \beta,\ov\delta}R^{A\eta}_{\ \ \ ;\alpha}) \nonumber\\
   &=& - {1\over y(v)}R^A_{\ \beta,\ov\delta}R^A_{\ov\sigma,\alpha}.
\end{eqnarray}
The second equality is due to 
\begin{eqnarray}
 R^A_{\ \beta}R^{A\eta} = y(v)\delta^\eta_\beta,
\end{eqnarray}
following from (4.4). Multiplying the Lie-algebra (2.14) by $R^A_{\ \gamma}$ or 
$R^{A\gamma}$ yields
\begin{eqnarray}
 R^{B\beta}_{\ \ \ ;\gamma} = {1\over y(v)}f^{ABC}R^{C\beta}R^A_{\ \gamma}
\end{eqnarray}
owing to (4.7), or 
\begin{eqnarray}
 -R^{B\alpha}R^{A\gamma}R^{A\beta}_{\ \ \ ,\alpha} 
   = f^{ABC}R^{C\beta}R^{A \gamma}.
\end{eqnarray}
The former is rewritten as
\begin{eqnarray}
 R^B_{\ov \beta,\gamma} = {1\over y(v)}f^{ABC}R^C_{\ \ov\beta}R^A_{\ \gamma},
\end{eqnarray}
while the latter becomes
\begin{eqnarray}
 f^{ABC}R^{C\beta}R^{A \gamma} = 0,
\end{eqnarray}
because we have
\begin{eqnarray}
 R^{A\gamma}R^{A\beta}_{\ \ \ ,\alpha} & =& R^{A\gamma}R^{A\beta}_{\ \ \ ;\alpha}
  \ \ =\ \ g^{\beta\ov \eta}R^{A\gamma}R^A_{\ \ov \eta,\alpha} \nonumber\\
  &=& -g^{\beta\ov \eta}R^{A\gamma}R^A_{\  \alpha,\ov \eta} \ \ =\ \ 0,
  \nonumber
\end{eqnarray}
due to (4.6), (2.15) and (4.9). With (4.12) the Riemann curvature (4.8) takes the form 
\begin{eqnarray}
 R_{\alpha\ov\sigma\beta\ov\delta} &=& {1\over y(v)^3} f^{ABE}R^A_{\  \alpha}R^B_{\ \ov 
\sigma}\cdot f^{CDE}R^C_{\  \beta}R^D_{\ \ov \delta}  \\
 &=& R_{\beta\ov\sigma\alpha\ov\delta}. \nonumber
\end{eqnarray}
The last equality follows from the symmetry of the Affine connection (4.7), or directly 
from the Jacobi identity of the structure constants
\begin{eqnarray}
-f^{ADC}f^{BCE} + f^{BDC}f^{ACE} = f^{ABC}f^{CDE},
\end{eqnarray}
and (4.13).
Contrary to (4.1) this manifests the isometry $G$ and includes no derivative with  respect 
to the coordinates. By using it the Riemann curvature can be calculated algebraically, once 
given a concrete form of the Killing vectors $R^{A\alpha}$  which are proper to the 
K\"ahler manifold $G/H$. Thus (4.14) gives a more practical formula than (4.1) for physical 
applications.

\subsection{The reducible case}

When the K\"ahler manifold $G/H$ is reducible, the broken generators $X^i$ are decomposed 
into irreducible sets under the subgroup $H$. Each irreducible set has a different 
$Y$-charge. The metric (3.15) satisfies the initial condition

\newfont{\bg}{cmr10 scaled \magstep4}
\newcommand{\bigzerol}{\smash{\hbox{\bg 0}}}
\newcommand{\bigzerou}{%
  \smash{\lower1.7ex\hbox{\bg 0}}}

\begin{eqnarray}
g_{\alpha\ov \beta}\mathop{\vert}_{\phi = \bar \phi = 0} & =&   
g_{\ov \beta\alpha}\mathop{\vert}_{\phi = \bar \phi = 0}    \nonumber\\ 
 &= & 
\left(
\begin{array}{ccccc}
y_1(v) &  &  &  &  \bigzerou \\ 
    & y_2(v) &  &  &   \\  
    &  & \cdot &  &  \\
    &  &  & \cdot &  \\
\bigzerol & & & & y_N(v) 
\end{array}\right).
\end{eqnarray}
Therefore the formula (4.4) is no longer correct in this case. We have to generalize the 
whole arguments in the previous subsection.

First of all, with $U$ given by (3.12) and a real symmetric matrix $P$ we define the 
quantity 
\begin{eqnarray}
 \Delta = UPU^{-1} \nonumber
\end{eqnarray}
in the adjoint representation of the isometry group $G$.
By (3.17) it transforms as
\begin{eqnarray}
 \Delta (\phi' \bar \phi') = g\Delta(\phi,\bar \phi)g^{-1}, \nonumber
\end{eqnarray}
or equivalently  
\begin{eqnarray}
 {\cal L_{R^A}}\Delta = i[T^A,\ \Delta],
\end{eqnarray}
 if $P$ satisfies  
\begin{eqnarray}
 hPh^{-1} = P.
\end{eqnarray}
With this $\Delta$ the metric $g^{ab}$ is found as a solution to 
  the Killing condition (2.7) 
\begin{eqnarray}
g^{ab} & = & g^{ba} \ \ = \ \ R^{Aa}\Delta^{AB}R^{Bb} \ \ \equiv \ \ R^a \Delta R^b .
\end{eqnarray}
In the complex basis it reads
\begin{eqnarray}
g^{\alpha\ov\beta}&=& g^{\ov\beta\alpha} \ \ = \ \ R^\alpha\Delta R^{\ov\beta},  
\nonumber\\
g^{\alpha\beta} &=&  R^\alpha\Delta R^{\beta},  \\
g^{\ov\alpha\ov\beta} &=& R^{\ov\alpha}\Delta R^{\ov\beta}. \nonumber   
\end{eqnarray}
 We now assume the real symmetric matrix $P$ to have non-vanishing elements only in the  
diagonal blocks corresponding to the broken generators $X^a = (X^{\ov i},X^i )$ such that 
\begin{eqnarray}
 P^{i\ov j} = P^{\ov j i} =
\left(
\begin{array}{ccccc}
y_1(v)^{-1} &  &  &  &  \bigzerou \\ 
    & y_2(v)^{-1} &  &  &   \\  
    &  & \cdot &  &  \\
    &  &  & \cdot &  \\
\bigzerol & & & & y_N(v)^{-1} 
\end{array}\right).
\end{eqnarray}
Then $P$ satisfies (4.18) because  the diagonal elements are decomposed into irreducible 
sets under the subgroup $H$ by the $Y$-charge.
Evaluate these metrics in (4.20) at the origin of the coset space by (4.5) and (4.21). We 
find that they all satisfy 
the same initial conditions as the metrics given in (3.15). Thus both metrics are 
equivalent\footnote{The equivalence is alternatively stated as 
$$
e^a = (R_\alpha UP)^ad\phi^\alpha + (R_{\ov\alpha} UP)^a d\phi^{\ov\alpha},
$$
with $e^a = (e^{\ov i}, e^i )$ defined by the Cartan-Maurer $1$-form (3.14). Namely both 
sides have the same Lie-derivatives with respect to $R^{Aa}$ and the same values at the 
origin of the coset space. The author is indebted to T. Kugo for the discusion on this 
comment }, and we have 
\begin{eqnarray}
 R^\alpha\Delta R^{\beta} = 0, \quad\quad R^{\ov\alpha}\Delta R^{\ov\beta} = 0.
\end{eqnarray}

This generalization of the metric requires to modify the formula (4.7)$\sim$ \\ (4.14) in 
the previous subsection. Rewrite (4.20) as 
\begin{eqnarray}
  R^\alpha \Delta R_\beta &=& \delta^\alpha_\beta, \quad\quad 
 R^{\ov \alpha} \Delta R_{\ov\beta} \ \ =\ \ \delta^{\ov\alpha}_{\ov\beta},\nonumber\\
 R^\alpha \Delta R_{\ov\beta} &=& 0, 
 \quad\quad\quad\ \  
 R^{\ov \alpha} \Delta R_{\beta} \ \ =\ \ 0, 
\end{eqnarray}
using (4.22).
Differentiate them by the coordinates to find 
\begin{eqnarray}
R_{\ov\alpha} (\Delta R_\beta)_{,\ov\gamma} \ \ =\ \ 0, \quad\quad \quad
 R_\alpha (\Delta R_{\ov\beta})_{,\gamma} &=& 0,  \\
 R_{\ov \alpha}( \Delta R_{\ov\beta})_{,\ov\gamma} \ \ =\ \ 0, \quad\quad\quad  
 R_\alpha( \Delta R_{\beta})_{,\gamma}& = &0.
\end{eqnarray}
With the metric (4.20)  the Affine connection (4.7) changes the form as
\begin{eqnarray}
\Gamma_{\alpha\beta}^{\ \ \gamma} &=& R_{\alpha,\beta}\Delta R^\gamma
                                  = - R_\alpha(\Delta R^\gamma)_{,\beta}\ , \nonumber
\end{eqnarray}
owing to (4.23) and (4.24). Then the Riemann curvature becomes 
\begin{eqnarray}
 R_{\alpha\ \beta\ov\delta}^{\ \gamma}
    & =& \Gamma_{\alpha\beta,\ov\delta}^{\ \ \gamma}  \nonumber\\
    & = & -[R_{\alpha,\ov\delta}(\Delta R^\gamma)_{,\beta} +
          R_\alpha(\Delta R^\gamma)_{,\beta\ov\delta}]  \nonumber\\
    &=& -[(R_{\alpha}\Delta)_{,\ov\delta} R^\gamma_{\  ,\beta} 
       + (R_\alpha \Delta_{,\beta})_{,\ov\delta}R^\gamma] .
\end{eqnarray}
By (4.24) and (2.15) the first piece changes the form as
\begin{eqnarray}
 (R_{\alpha}\Delta)_{,\ov\delta} R^\gamma_{\ ,\beta}  &=&
     (R_{\alpha}\Delta)_{,\ov\delta} R^\gamma_{\  ;\beta}   \nonumber\\
    &=& g^{\gamma\ov\sigma}(R_\alpha\Delta)_{,\ov\delta}R_{\ov\sigma,\beta}
       \nonumber\\
     &=& - g^{\gamma\ov\sigma}(R_\alpha\Delta)_{,\ov\delta}R_{\beta,\ov\sigma}.
     \nonumber
\end{eqnarray}
By means of the formulae (A.7) in the Appendix A it becomes
\begin{eqnarray}
 (R_{\alpha}\Delta)_{,\ov\delta} R^\gamma_{\ ,\beta}  &=&
   g^{\gamma\ov\sigma}\{(R_\alpha \Delta)_{,\ov\delta} R^{\ov\eta}
     \cdot R_{\ov\sigma} \Delta_{,\ov\eta} R_{\beta}  \nonumber\\
    & + &  f^{ABC}(R_\alpha \Delta)^A_{\ ,\ov\delta}
        (R_{\ov\sigma}\Delta)^B R_{\ \beta}^C \}.
\end{eqnarray}
On the other hand the second piece of (4.26) is calculated as
\begin{eqnarray}
 (R_\alpha \Delta_{,\beta})_{,\ov\delta}R^\gamma
   &=&  g^{\gamma\ov\sigma} \{ f^{ABC}(R_\alpha \Delta)^A_{\ \ ,\ov\delta}
       R^B_{\  \beta}
      ( R_{\ov\sigma}\Delta)^C   \nonumber\\
   & & \hspace{0.5cm}+  f^{ABC}(R_\alpha \Delta)^A
       R^B_{\  \beta,\ov\delta}
       (R_{\ov\sigma}\Delta)^C   \\
   & &\hspace{0.5cm}  + f^{ABC}(R_\alpha \Delta)^A
       R^B_{\  \beta}
       (R_{\ov\sigma}\Delta_{,\ov\delta})^C\},   \nonumber
\end{eqnarray}
by means of (A.10). Putting (4.27) and (4.28) together into (4.26) we have 
\begin{eqnarray}
 R_{\alpha\ov\sigma\beta\ov\delta} & = & 
 -(R_\alpha \Delta)_{,\ov\delta} R^{\ov\eta}
     \cdot R_{\ov\sigma} \Delta_{,\ov\eta} R_{\beta} \nonumber\\
  &-&   f^{ABC}(R_\alpha \Delta)^A
       R^B_{\  \beta,\ov\delta}
       (R_{\ov\sigma}\Delta)^C   \\
   &-&  f^{ABC}(R_\alpha \Delta)^A
       R^B_{\  \beta}
       (R_{\ov\sigma}\Delta_{,\ov\delta})^C .  \nonumber
\end{eqnarray}
Calculate the first piece further as
\begin{eqnarray}
  (R_\alpha \Delta)_{,\ov\delta} R^{\ov\eta}  
    \cdot  R_{\ov\sigma} \Delta_{,\ov\eta} R_{\beta}  
     &=& (R_\alpha \Delta)_{,\ov\delta} R_\rho \cdot g^{\rho\ov\eta}\cdot
     R_{\ov\sigma} \Delta_{,\ov\eta} R_{\beta} \nonumber\\
     &=& R_\alpha \Delta R_{\ov\delta,\rho} \cdot g^{\rho\ov\eta}\cdot
     R_{\ov\sigma} \Delta_{,\ov\eta} R_{\beta} \nonumber\\ 
     &=& - R_\alpha \Delta_{,\rho}R_{\ov\delta}\cdot g^{\rho\ov\eta}\cdot
     R_{\ov\sigma} \Delta_{,\ov\eta} R_{\beta}, 
\end{eqnarray}
in which the second equality is obtained by (4.22) and (2.15), while the third one by 
(4.24). Rewrite the last piece of (4.29) as  
\begin{eqnarray}
f^{ABC}(R_\alpha \Delta)^A
       R^B_{\  \beta}
       (R_{\ov\sigma}\Delta_{,\ov\delta})^C  &=& 
    R_\alpha \Delta_{,\beta} R_{\ov\eta}\cdot 
   R_{\ov\sigma} \Delta_{,\ov\delta} R^{\ov \eta}, 
\end{eqnarray}
 by (A.8).
By means of (A.7), (A.9) and (A.10) the Riemann curvature (4.29) turns out to be 
\begin{eqnarray}
 R_{\alpha\ov\sigma\beta\ov\delta} 
  &=& \ f^{ABC}(R_\alpha \Delta)^A(R_{\ov\delta}\Delta)^B R^{C\ov\eta}\cdot
      f^{DEF}(R_{\ov\sigma}\Delta)^D (R_\beta \Delta)^E R_{\ \ov\eta}^F
        \nonumber \\
 &+ & \ f^{ABC}(R_\alpha \Delta)^A(R_{\ov\sigma}\Delta)^B R^{C\ov\eta}\cdot
      f^{DEF}(R_{\ov\delta} \Delta)^D (R_\beta \Delta)^E R_{\ \ov\eta}^F
         \nonumber \\
 &+&   f^{ABC}(R_\alpha \Delta)^A(R_{\ov\sigma}\Delta)^B \cdot
      f^{CDE}R^D_{\ \beta}( R_{\ov\delta}\Delta)^E
      \nonumber\\
 & -& \ f^{ABC}(R_\alpha \Delta)^A R^B_{\ \beta} (R_{\ov \eta}\Delta)^C \cdot
   f^{DEF}(R_{\ov\sigma}\Delta)^D R^E_{\ \ov\delta} (R^{\ov\eta}\Delta)^F.    
\end{eqnarray}
This is the generalization of (4.14). 
If one replaces $\Delta^{AB}$ by ${1\over y(v)}\delta^{AB}$, then 
(4.32) reduces to (4.14) due to the formula (4.13) which is only valid for the irreducible 
case. At this final stage it is worth showing the symmetry property 
\begin{eqnarray}
 R_{\alpha\ov\sigma\beta\ov\delta} &=& R_{\beta\ov\sigma\alpha\ov\delta} \ \ = \ \ 
R_{\alpha\ov\delta\beta\ov\sigma},
\end{eqnarray}
 as a consistency check of our calculations. The demonstration will be given in Appendix B. 
There we also show that the formula (4.32) takes an alternative form such that
\begin{eqnarray}
R_{\alpha\ov\sigma\beta\ov\delta} 
  &=& \ f^{ABC}(R_\alpha \Delta)^A(R_{\ov\delta}\Delta)^B R^{C\ov\eta}\cdot
      f^{DEF}(R_{\ov\sigma}\Delta)^D (R_\beta \Delta)^E R_{\ \ov\eta}^F
        \nonumber \\
 &+ & \ f^{ABC}(R_\beta \Delta)^A(R_{\ov\delta} \Delta)^B  R^{C \ov\eta}\cdot 
      f^{DEF}(R_{\ov\sigma}\Delta)^D(R_\alpha \Delta)^E R_{\ \ov\eta}^F
    \nonumber \\
 &+ &  f^{ABC}(R_\alpha \Delta)^A(R_{\ov\sigma}\Delta)^B \cdot
      f^{CDE}(R_{\ \beta}\Delta)^D R^E_{\ \ov\delta}
      \nonumber\\
 & -& \ f^{ABC}(R_\alpha \Delta)^A R^B_{\ \beta} (R_{\ov \eta}\Delta)^C \cdot
   f^{DEF}(R_{\ov\sigma}\Delta)^D R^E_{\ \ov\delta} (R^{\ov\eta}\Delta)^F. 
   \nonumber
\end{eqnarray}

\section{Applications}
\setcounter{equation}{0}

In the $N=1$ supersymmetric non-linear $\sigma$-model on the K\"ahler coset space $G/H$, 
the four-fermi coupling is the most interesting part.  When the coset space is reducible, 
the Riemann curvature depends  on the $Y$-charge of the broken generators through the 
metric (3.15). It takes the form (4.32) which is rather complicated than that of the 
irreducible coset space.  On top of this complication we have another one, if the 
homogeneous subgroup $H$ contains plural $U(1)$s as $H = H'\otimes \{ U(1)\}^k$. Namely,
the splitting of the broken generators $X^{\ov i}$ and $X^i$ depends on 
 the constants $v^\mu$ of the $Y$-charge (3.2), so that we may have different sets of the 
NG bosons\cite{8} . Of course a phenomelogically interesting set should be chosen. Then the 
four-fermi coupling depends on the $Y$-charges of the broken generators $X^{\ov i}$ in a 
piculiar way to the choice. It is very interesting from the phenomelogical point of view.

As has been explained in the introduction,
the most important part of the four-fermi coupling is
\begin{eqnarray}
R_{\alpha\ov\sigma\beta\ov\delta}\mathop{\vert}_{\phi = \bar \phi = 0}
  (\bar \psi^{\ov\sigma}\psi^\alpha)(\bar\psi^{\ov\delta}\psi^\beta)
\end{eqnarray}
in the non-linear $\sigma$-model as a low-energy effective theory. We shall present a 
systematic method to evaluate the four-fermi coupling constants
$R_{\alpha\ov\sigma\beta\ov\delta}\mathop{\vert}_{\phi = \bar \phi = 0}$
 by means of (4.32). The method will enable us to fully controle the $Y$-charge dependence 
of $R_{\alpha\ov\sigma\beta\ov\delta}\mathop{\vert}_{\phi = \bar \phi = 0}$.

By (4.5) and (4.21) we note at first that 
\begin{eqnarray}
R_{\ \alpha}^A \mathop{\vert}_{\phi = \bar \phi = 0} &=& -iy_\alpha(v)\delta^A_\alpha,   
\quad\quad\quad    
R_{\ \ov\alpha}^A \mathop{\vert}_{\phi = \bar \phi = 0}\ \  = \ \ 
iy_\alpha(v)\delta^A_{\ov\alpha}.  \nonumber  
\end{eqnarray}
and
\begin{eqnarray}
(R_\alpha \Delta)^A \mathop{\vert}_{\phi = \bar \phi = 0} &=& -i\delta^A_\alpha,      \quad 
\quad\quad  (R^{\ov\alpha}\Delta)^A \ \ = \ \  -{i\over y_\alpha(v)}\delta^{\ov\alpha A}   
\nonumber\\
(R_{\ov\alpha} \Delta)^A \mathop{\vert}_{\phi = \bar \phi = 0} &=& i\delta^A_{\ov\alpha},      
\quad \quad\quad  (R^\alpha \Delta)^A \ \ =\ \ {i\over y_\alpha(v)}\delta^{\alpha A}   
\nonumber
\end{eqnarray}
By using this (4.32) becomes
\begin{eqnarray}
R_{\alpha\ov\sigma\beta\ov\delta}\mathop{\vert}_{\phi = \bar \phi = 0}
 &=& -\sum_{\eta = 1}^N 
   (f^{\ov\alpha\delta\ov\eta}\cdot f^{\ov\beta\sigma\eta}
    + f^{\ov\alpha\sigma\ov\eta}\cdot f^{\ov\beta\delta\eta})y_{\eta}(v)
      \nonumber\\
  &+ & [\sum_{\eta = 1}^N 
   (f^{\ov\alpha\sigma\eta}\cdot f^{\ov\beta\delta\ov\eta}
    + f^{\ov\alpha\sigma\ov\eta}\cdot f^{\ov\beta\delta\eta})y_{\beta}(v)
   + \sum_{C= I,\mu}^{{\rm dim}H} 
   f^{\ov\alpha\sigma C}\cdot f^{\ov\beta\delta C}y_{\beta}(v)] \nonumber\\
  &-& \sum_{\eta = 1}^N f^{\ov\alpha\ov\beta\eta}\cdot f^{\sigma\delta\ov\eta} 
  \ {y_\beta(v) y_\delta(v) \over y_\eta(v)}.  
\end{eqnarray}
Each piece of (5.2) can be computed by means of 
\begin{eqnarray}
f^{ABC}f^{CDE} &=& -{1\over 2{\cal N}} {\rm tr}([T^A,\ T^B][T^D,\ T^E]),  
\end{eqnarray}
with the normalization $tr(T^AT^B)=2{\cal N}\delta^{AB}$. The Riemann curvature appears 
with the indices  $\alpha,\ov\sigma,\beta,\ov\delta$ of the three types:
\begin{eqnarray}
y([T^{\ov\alpha},\ T^\sigma ] ) &>& 0, \quad \quad \quad 
y([T^{\ov\beta},\ T^\delta ] ) \ \ < \ \  0 ,  \\
y([T^{\ov\alpha},\ T^\sigma ] ) &<& 0, \quad \quad \quad 
y([T^{\ov\beta},\ T^\delta ] ) \ \ > \ \  0 ,  \\
y([T^{\ov\alpha},\ T^\sigma ] ) &=& 0, \quad \quad \quad 
y([T^{\ov\beta},\ T^\delta ] )\ \  = \ \  0 ,  
\end{eqnarray}
in which $y([\ ,\ ])$ is the $Y$-charge of the commutator.
(4.32) reads 
\begin{eqnarray}
R_{\alpha\ov\sigma\beta\ov\delta}\mathop{\vert}_{\phi =  \bar \phi = 0}
  &=& \sum_{\eta = 1}^N 
   f^{\ov\alpha\sigma\eta}\cdot f^{\ov\beta\delta\ov\eta}\ y_{\beta}(v)
   \nonumber\\
  &-&\sum_{\eta = 1}^N
  [f^{\ov\alpha\delta\ov\eta}\cdot f^{\ov\beta\sigma\eta}\ y_{\eta}(v)
     \ \ + \ \ f^{\ov\alpha\ov\beta\eta}\cdot f^{\sigma\delta\ov\eta} 
  \ {y_\beta(v) y_\delta(v) \over y_\eta(v)}]
\end{eqnarray}
for the case (5.4), 
\begin{eqnarray}
R_{\alpha\ov\sigma\beta\ov\delta}\mathop{\vert}_{\phi = \bar \phi = 0}
  &=& \sum_{\eta = 1}^N 
   f^{\ov\alpha\sigma\ov\eta}\cdot f^{\ov\beta\delta\eta}\ (y_{\beta}(v) 
  \ \ -\ \   y_{\eta}(v))
    \nonumber\\
 &-&\sum_{\eta = 1}^N
 [f^{\ov\alpha\delta\ov\eta}\cdot f^{\ov\beta\sigma\eta}y_\eta(v)
   \ \ +\ \ f^{\ov\alpha\ov\beta\eta}\cdot f^{\sigma\delta\ov\eta} 
  \ {y_\beta(v) y_\delta(v) \over y_\eta(v)}]
\end{eqnarray}
for the case (5.5), and
\begin{eqnarray}
R_{\alpha\ov\sigma\beta\ov\delta}\mathop{\vert}_{\phi = \bar \phi = 0}
  &=&  \sum_{C= I,\mu}^{{\rm dim}H} 
   f^{\ov\alpha\sigma C}\cdot f^{\ov\beta\delta C}\ y_{\beta}(v)  \nonumber\\
  &-& \sum_{\eta = 1}^N [f^{\ov\alpha\delta\ov\eta}\cdot f^{\ov\beta\sigma\eta}\ 
y_{\eta}(v) 
  + f^{\ov\alpha\ov\beta\eta}\cdot f^{\sigma\delta\ov\eta} 
  \ {y_\beta(v) y_\delta(v) \over y_\eta(v)}]
\end{eqnarray}
for the case (5.6). (5.7) can be alternatively obtained by applying the symmetry property 
(4.33) to (5.8).

\subsection{$SU(3)/\{U(1)\}^2$}

We start with the most simplest reducible coset space $SU(3)/\{U(1)\}^2$ to illustrate our 
basic strategy. The generators of $SU(3)$ are
\begin{eqnarray}
\{T^A\} &=& \{T^1_2,\ T^1_3,\ T^2_3,\ T^2_1,\ T^3_1,\ T^3_2,\ Q,\ Q' \}, \nonumber
\end{eqnarray}
with
\begin{eqnarray} 
Q = \textstyle{1\over \sqrt 2}(T^1_1 - T^2_2), \quad\quad\quad 
Q' = -\textstyle{\sqrt {3\over 2}}T^3_3,
\end{eqnarray}
and the hermitian  condition $(T^i_j)^\dagger = T^i_j$. They satisfy the Lie-algebra
\begin{eqnarray}
[T^j_i,\ T^l_k] = \delta_k^j T^l_i - \delta^l_i T^j_k. \nonumber
\end{eqnarray}
The quadratic Casimir takes the form
\begin{eqnarray}
\{T^1_2,T^2_1\} + \{T^1_3,T^3_1\}+ \{T^2_3,T^3_2\} + Q^2 + Q'^2 \ , \nonumber
\end{eqnarray}
from which we read the Killing metric $\delta^{AB}$ in (2.18).
The $U(1)$-charges $Q$ and $Q'$ of the broken generators $T^j_i \ (i\ne j)$
as well as their $Y$-charges  
\begin{eqnarray}
Y & = & vQ + v'Q'  \nonumber
\end{eqnarray}
 are  given in Table 1.
By means of them  the broken generators  are splitted in 

\begin{table}[h]
\begin{center}
\begin{tabular}{|c|r|r|c|}
\hline
$X^{\ov i}$ & $Q\quad $ & $Q'\quad $ & $y(X^{\ov i})$  \\ \hline
 $T^1_2$ &$-\sqrt 2\quad $ &$0\quad $
  & $-\sqrt 2 v$  \\ \hline
 $T^1_3$ &$-{1\over \sqrt2}\quad $ &$-\sqrt{3\over 2 }\quad $
  & $-{1\over \sqrt2}v -\sqrt{3\over 2}v'$ \\ \hline
  $T^2_3$ &${1\over \sqrt2}\quad $ &$-\sqrt{3\over 2}\quad $ 
  &${1\over \sqrt2}v -\sqrt{3\over 2}v'$  \\ 
  \hline
\end{tabular}
\caption{$U(1)$-charges of $X^{\ov i}$ in $SU(3)$.}
\end{center}  
\end{table}

\noindent
 two parts: the generators $X^i$ with positive $Y$-charge and their hermitian conjugates 
$X^{\ov i}$ with negative charge. For illustration we plot the broken generators in the 
$(Q,Q')$-charge plane in Figure 1. 
There are three possibilities to draw the line : $Y = 0$,

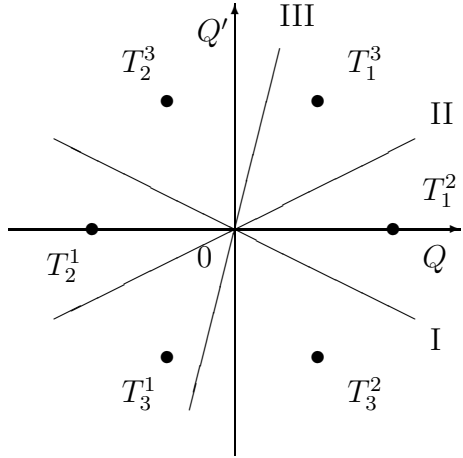
\begin{figure}[hbt]
\setlength{\unitlength}{1mm}
\begin{picture}(60,60)(-30,-60)

\put(0,-30){\vector(1,0){60}}
\put(30,-60){\vector(0,1){60}}
\put(50,-31){$\bullet$}
\put(55,-26){$T^2_1$}
\put(10,-31){$\bullet$}
\put(5,-36){$T^1_2$}
\put(40,-14){$\bullet$}
\put(45,-9){$T^3_1$}
\put(40,-48){$\bullet$}
\put(45,-53){$T^2_3$}
\put(20,-14){$\bullet$}
\put(15,-9){$T^3_2$}
\put(20,-48){$\bullet$}
\put(15,-53){$T^1_3$}

\put(25,-35){$0$}
\put(55,-35){$Q$}
\put(25,-5){$Q'$}

\put(54,-18){\line(-2,-1){48}}
\put(56,-16){II}
\put(54,-42){\line(-2,1){48}}
\put(56,-46){I}
\put(36,-6){\line(-1,-4){12}}
\put(36,-3){III}

\end{picture}
\caption{The splitting of the broken generators}

\end{figure}

\noindent
 each of which gives a different splitting:
\begin{eqnarray}
{\rm I} \hspace{1.5cm} \{X^i\} &=& \{T^2_1,\ T^3_1,\ T^3_2\},  \nonumber \\
{\rm II}\hspace{1.5cm} \{X^i\} &=& \{T^3_1,\ T^3_2,\ T^1_2 \},  \nonumber \\
{\rm III} \hspace{1.5cm} \{X^i\} &=& \{T^3_2,\ T^1_2,\ T^1_3 \}.  \nonumber 
\end{eqnarray}

Taking the case I we proceed with the argument. With the identification $X^1 = 
T^2_1,\ X^2 = T^3_1, \ X^3 = T^3_2,\ Q=X^q,\ Q'=X^{q'}$, the non-trivial part of the 
Lie-algebra  reads 
\begin{eqnarray}
[X^2,\ X^{\ov 3}] &=& X^1, \quad\quad\quad [X^3,\ X^{\ov 2}]\ \ = \ \ X^{\ov 1},
   \nonumber \\     \nonumber  
[X^1,\ X^3] & = & X^2, \quad\quad\quad [X^{\ov 3},\ X^{\ov 1}] \ \ = \ \ X^{\ov 2},  \\ 
\nonumber
[X^{\ov 1}, \ X^2] &=&  X^3, \quad\quad\quad [X^{\ov 2},\ X^1]\ \ = \ \ X^{\ov 3},   \\  
\nonumber
[X^1, \ X^{\ov 1}] &=& \sqrt 2 Q, \\ \hspace{0cm}
[X^2, \ X^{\ov 2}] &=& \textstyle{\frac1{\sqrt2}}Q +\textstyle{\sqrt{\frac3 2}}Q', \\ 
\nonumber
[X^3, \ X^{\ov 3}] &=& -\textstyle{\frac1{\sqrt2}}Q +\textstyle{\sqrt{\frac3 2}}Q', \\ 
\nonumber
[X^q,\ X^1] &=& \sqrt2 X^1,\quad\quad\quad [X^{q'},\ X^1]  = 0, 
  \\ \nonumber
[X^q,\ X^2] &=& \textstyle{1\over \sqrt2}X^2,
\quad\quad [X^{q'},\ X^2]  =  \textstyle\sqrt{3\over 2}X^2,
  \\ \hspace{0cm}
[X^q,\ X^3] &=& -\textstyle{1\over \sqrt2}X^3,
\ \quad [X^{q'},\ X^3]  = \textstyle\sqrt{3\over 2}X^3,\nonumber
 \end{eqnarray}
and their hermitian conjugates. 
The holomorphic Killing vectors $R^{A\alpha}$ are easily obtained by studying (3.6) in the 
fundamental representation:
$$
\begin{array}{lll}
R^{\ov 1 1} = i,\hspace{0.5cm} & R^{11} = -i(\phi^1)^2 \hspace{0.5cm}& 
 R^{q1}= -\sqrt2 i\phi^1,   \\
 R^{\ov 2 1}=0,\hspace{0.5cm} & R^{21}= -i\phi^1(\phi^2 + {1\over 
2}\phi^1\phi^3),\hspace{0.5cm} &  R^{q'1} = 0,     \\
  R^{\ov 3 1}= 0, \hspace{0.5cm}&  R^{31} = i(\phi^2 + {1\over 2}\phi^1\phi^3), &   \\
R^{\ov 1 2} = -{i\over 2}\phi^3,\hspace{0.5cm} & R^{12} = -{i\over 2}\phi^1(\phi^2 + 
{1\over 2}\phi^1\phi^3), \hspace{0.5cm}& 
 R^{q2}= -{i\over \sqrt2}\phi^2,   \\
 R^{\ov 2 2}= i,\hspace{0.5cm} & R^{22}=-i[(\phi^2)^2 + {1\over 
4}(\phi^1\phi^3)^2],\hspace{0.5cm} &  R^{q'2} = -\textstyle\sqrt{3\over 2}i\phi^2,     \\
  R^{\ov 3 2}={i\over \sqrt2}\phi^1 , \hspace{0.5cm}&  R^{32} = -{i\over 2}\phi^1(\phi^2 - 
{1\over 2}\phi^1\phi^3), &   \\
R^{\ov 1 3} = 0,\hspace{0.5cm} & R^{13} = -i(\phi^2 - {1\over 2}\phi^1\phi^3), 
\hspace{0.5cm}& 
 R^{q3}= {i\over \sqrt2}\phi^3,   \\
 R^{\ov 2 3}=0,\hspace{0.5cm} & R^{23}= -i\phi^1(\phi^2 - {1\over 
2}\phi^1\phi^3),\hspace{0.5cm} &  R^{q'3} = -\textstyle\sqrt{3\over 2}i\phi^3,     \\
  R^{\ov 3 3}=i , \hspace{0.5cm}&  R^{33} = -i(\phi^3)^2, &   \\ 
\end{array}
$$
The Riemann curvature $R_{\alpha\ov\sigma\beta\ov\delta}\mathop{\vert}_{\phi = \bar \phi = 
0} (\equiv G^{\ov\alpha\sigma\ov\beta\delta}$), given by
(5.2), is calculated by using (5.3) with the commutators (5.11). 
The Riemann curvature of this type  appears in the coset space $E_7/SU(5)\otimes 
\{U(1)\}^3$ which we will study in the next subsection. The result is  given in Tables 8 
and 9 there.

\subsection{$E_7/SU(5)\otimes \{U(1)\}^3$}

The generators of $E_7$ are decomposed as 
\begin{eqnarray}
\{T^A\} = \{ E^{ab}_i,\ T^i_a,\ E^a,\ E^i_{ab},\ T^a_i,\  E_a, \ T_a^b,\ T^j_i,\ T \}
\end{eqnarray}
in the basis of the subgroup $SU(5)\otimes SU(3)\otimes U(1)$. Here $a,b,\cdots$ and 
$i,j,\cdots$ are indices of $SU(5)$ and $SU(3)$ running over $1\sim 5$ and $1\sim 3$ 
respectively. They have $SU(5)\otimes SU(3)$ quantum numbers
 \begin{eqnarray}
({\bf 10},{\bf 3^*}), ({\bf 5^*},{\bf 3}),({\bf 5},{\bf 1}),({\bf 10^*},{\bf 3}), ({\bf 5},
{\bf 3^*}),({\bf 5^*},{\bf 1}),({\bf 24},{\bf 1}),({\bf 1},{\bf 8}),({\bf 1},{\bf 1}), 
\nonumber
\end{eqnarray}
in the order of (5.12). The non-trivial part of the $E_7$ algebra takes the 
form\cite{8}\cite{15}  
\begin{eqnarray}
  \hspace{0cm} 
& &  [E^{ab}_i ,\ E^{cd}_j ] = \varepsilon_{ijk}\varepsilon^{abcde}
T^k_e,  \hspace{1cm} [E^a\ ,\ E^{bc}_j ] = 0,   \nonumber   \\
 \hspace{0cm} & & [T^i_a \ ,\ E^{bc}_j \ ] = \delta^i_j (\delta^b_a E^c - \delta^c_a E^b),   
\hspace{1cm}  [T^i_a \ ,\ E^b \ ] = 0,  \nonumber\\
 \hspace{0cm} & & [E^a\ ,\ E^b\ ] = 0,      \nonumber\\
 \hspace{0cm} & & [E_{ab}^i,\ E^{cd}_j ] = \delta^i_j(\delta^c_a T^d_b + 
 \delta^d_b T^c_a - \delta^c_b T^d_a - \delta^d_a T^c_b )
   - \delta^{cd}_{ab}(T^i_j + \textstyle\sqrt{2\over 15}\delta^i_j T), 
   \nonumber\\
 \hspace{0cm}& & [E^a\ , \ E_b \ ] = \textstyle\sqrt{6\over 5}\delta^a_b T - T^a_b,  \\
 \hspace{0cm}& & [E^a\ ,\ E^i_{bc}] = \delta^a_cT^i_b - \delta^a_bT^i_c, 
 \nonumber\\
\hspace{0cm} & & [T^i_a\ ,\ T^b_j \ ] = -\delta_a^bT^i_j + \delta^i_jT^b_a 
  +2\textstyle\sqrt{2\over 15}\delta^i_j\delta^b_a T,  \nonumber\\
\hspace{0cm}& & [T^i_a\ , \ E^j_{bc}] = -\textstyle{1\over 
2}\varepsilon^{ijk}\varepsilon_{abcde}E^{de}_k,  \nonumber\\
\hspace{0cm}& & [T^i_a\ ,\ E_b\ ] = -E^i_{ab}, \quad\quad {\rm h.c.}, \nonumber
\end{eqnarray}
with
$$
 \varepsilon_{ijk}^* = \varepsilon^{ijk},\quad
\varepsilon_{abcde}^* = \varepsilon^{abcde}, \quad
 \delta^{cd}_{ab} = \delta^c_a\delta^d_b - \delta_a^d\delta^c_b.
$$
In the basis of the smaller subgroup  $SU(5)\otimes \{U(1)\}^3$ the generators are further 
decomposed as 
\begin{eqnarray}
\{T^A\} = \{X^{\ov i},\ X^i,\ S^I,\ Q^\mu \},
\end{eqnarray}
with 
\begin{eqnarray}
\{X^{\ov i}\}&=& \{E_i^{ab},\ E^a,\ T_a^i,\ T^j_i \ (i > j)\},  \nonumber\\
\{X^i\}&=& \{E^i_{ab},\ E_a,\ T^a_i,\ T^j_i \ (i < j)\},
    \nonumber\\
 \{S^I\} &=& \{T^b_a\ \ \ ( \textstyle{\sum_a} T^a_a = 0) \},  \\
 \{Q^\mu\} &=& \{T,\ Q,\ Q'\}.  \nonumber
\end{eqnarray}
Here $Q$ and$Q'$ are the $U(1)$s contained in $SU(3)$ which was given by (5.10). The 
quadratic Casimir takes the form 
\begin{eqnarray}
C &=& {1\over 2}\{E^{ab}_i,E_{ab}^i\} + \{T^a_i\ ,T^i_a\ \} + \{E^a\ ,E_a\ \} 
\nonumber \\
  &+ &   \{T_2^1\ ,T_1^2\ \}
 + \{T_3^1\ ,T_1^3\ \} + \{T_3^2\ ,T_2^3\ \}  \nonumber\\
 &+ &   \{T_a^b\ ,T_b^a\ \} + T^2 + Q^2 + Q'^2\ . \nonumber
\end{eqnarray}
$\{X^{\ov i}\}$ and $\{X^i\}$ are broken generators of $E_7/SU(5)\otimes \{U(1)\}^3$.  In 
the supersymmetric model on  $E_7/SU(5)\otimes \{U(1)\}^3$ there are pseud NG fermions 
corresponding to them. Among of them the pseud NG fermions having the same $SU(5)\otimes 
SU(3)$ quantum numbers as $E_i^{ab},\  T_a^i$ and $T_i^j (i>j)$ are identified with the 
three families of quarks and leptons $\psi^{ab \atop i}, \psi^{i\atop a}$ and $\psi^{j\atop 
i}$.
The $Y$-charge is made of the $U(1)$ charges in $\{ Q^\mu \}$:
\begin{eqnarray}
Y = \alpha T + \beta Q + \gamma Q'.
\end{eqnarray}
These $U(1)$-charges are given in Table 3.

\begin{table}[hbt]
\begin{center}
\begin{tabular}{|c|c|r|r|c|}
\hline
 $X^{\ov i}$ &$T$ & $Q\quad $ & $Q'\quad $ & $y(X^{\ov i})$  \\ \hline
$E^{ab}_1$ & $1$ & ${1\over \sqrt2}\quad   $ & ${1\over \sqrt6}\quad   $   
& $ \alpha + {1\over \sqrt2}\beta + {1\over \sqrt6}\gamma$   \\  \hline
$E^{ab}_2$ & $1$ & $-{1\over \sqrt2}\quad  $ & ${1\over \sqrt6}\quad $  
& $ \alpha - {1\over \sqrt2}\beta + {1\over \sqrt6}\gamma$    \\  \hline
$E^{ab}_3$ & $1$ & $0\quad $ & $-\sqrt{2\over 3}\quad $ 
 & $ \alpha -\sqrt{2\over 3}\gamma  $   \\  \hline
$T^1_a$  & $2$ & $-{1\over \sqrt2}\quad $ &$-{1\over \sqrt6}\quad $  
& $2\alpha-{1\over \sqrt2}\beta-{1\over \sqrt6}\gamma$    \\  \hline
$T^2_a $ & $2$ &${1\over \sqrt2}\quad $ & $-{1\over \sqrt6}\quad $ 
 & $2\alpha+{1\over \sqrt 2}\beta-{1\over \sqrt6}\gamma\quad $   \\  \hline
$T^3_a$  & $2$ &$0\quad $  & $\sqrt{2\over 3}\quad $
 &$2\alpha+\sqrt{2\over 3}\gamma$  \\  \hline
$E^a$  & $3$ &$0\quad $ & $0\quad $ 
& $3\alpha$ \\ \hline
 $T^1_2$ & $0$ &$-\sqrt 2\quad $ &$0\quad $
  & $-\sqrt 2\beta$  \\ \hline
 $T^1_3$ & $0$ &$-{1\over \sqrt2}\quad $ &$-\sqrt{3\over 2 }\quad $
  & $-{1\over \sqrt2}\beta -\sqrt{3\over 2}\gamma$ \\ \hline
  $T^2_3$ & $0$ &${1\over \sqrt2}\quad $ &$-\sqrt{3\over 2}\quad $ 
  &${1\over \sqrt2}\beta -\sqrt{3\over 2}\gamma$  \\ 
  \hline
\end{tabular}
\caption{$U(1)$-charges of $X^{\ov i}$ in $E^7$.}
\end{center}  
\end{table}

 The splitting of the broken generators $\{X^{\ov i}\}$ and $\{X^i\}$ changes according to 
the orientation of the plane $Y = 0$ in the $(T,Q,Q')$-charge space.
 The splitting (5.15) is valid only when  the vector coefficients $(\alpha,\beta,\gamma)$ 
are chosen such that
\begin{eqnarray}
y_i(v) \equiv  Y(X^{\ov i}) > 0 , 
   \quad\quad  {\rm for \ \  all}\ \ov i ,  \nonumber
\end{eqnarray}
for instance, 
\begin{eqnarray}
  \alpha = 1, \quad \beta < 0, \quad \gamma < 0 ,\quad |\beta| = |\gamma| << 1. \nonumber
\end{eqnarray}
We proceed with the argument in this special splitting, since the pseud NG fermions are 
then neatly identified with the three families of quarks and leptons.

In the supersymmetric $\sigma$-model on $E_7/SU(5)\otimes \{U(1)\}^3$ the Riemann curvature 
$R_{\alpha\ov\sigma\beta\ov\delta}$ is a $SU(5)$-covariant tensor. We shall be interested 
in the four-fermi coupling of the three families of quarks and leptons alone. Then the 
relevant part of the Riemann curvature appears with the $SU(5)$-content of the following 
types:
$$
R_{\alpha\ov\sigma\beta\ov\delta}\mathop{\vert}_{\phi = \bar \phi = 0}
 \equiv \ \ G^{\ov\alpha\sigma\ov\beta\delta}
 \sim \ \  \left\{    
\begin{array}{ll}
   ({\bf  5}^*,{\bf \ 5\ },{\bf \ 5}^*,{\bf \ 5\ }),&  \\
   ({\bf 10},{\bf 10}^*,{\bf 10},{\bf 10}^*),&  \\
   ({\bf 1},{\bf 1},{\bf 1},{\bf 1}), & \\
   ({\bf  5}^*,{\bf \ 5},{\bf 10},{\bf 10}^*),& 
   ({\bf 10},{\bf 10}^*,{\bf  5}^*,{\bf \ 5}),\\
   ({\bf 5},{\bf 10},{\bf  5}^*,{\bf \ 10}^*), &
   ({\bf  5}^*,{\bf \ 10}^*,{\bf 5},{\bf 10}),  \\
   ({\bf 1},{\bf 1},{\bf 5}^*,{\bf 5}), &
   ({\bf 5}^*,{\bf 5},{\bf 1},{\bf 1}), \\
   ({\bf 1},{\bf 5},{\bf 5}^*,{\bf 1}), &
   ({\bf 5}^*,{\bf 1},{\bf 5},{\bf 1}), \\
   ({\bf 1},{\bf 1},{\bf 10},{\bf 10}^*), &
   ({\bf 10},{\bf 10}^*,{\bf 1},{\bf 1}), \\
   ({\bf 1},{\bf 10}^*,{\bf 10},{\bf 1}), &    
   ({\bf 10},{\bf 1},{\bf 1},{\bf 10}^*).    
\end{array}   \right.
$$
We evaluate $G^{\ov\alpha\sigma\ov\beta\delta}$ in components by means of (5.2) with (5.3) 
and (5.13). (See Appendix C.)  The results are summarized in Tables 3$\sim$16 
   for six types of the Riemann curvature 
$$
\begin{array}{lll}
({\bf  5}^*,{\bf \ 5\ },{\bf \ 5}^*,{\bf \ 5\ }),& 
   ({\bf 10},{\bf 10}^*,{\bf 10},{\bf 10}^*),&
   ({\bf  1},{\bf \ 1},{\bf \ 1},{\bf \ 1}),  \\
    ({\bf  5}^*,{\bf \ 5},{\bf 10},{\bf 10}^*),  &
    ({\bf \ 1},{\bf \ 1},{\bf \ 5}^*,{\bf \ 5}),&
    ({\bf  1},{\bf \ 1},{\bf 10},{\bf 10}^*). 
\end{array}
$$
 Other types of  the Riemann curvature  are obtained by applying the symmetry property 
(4.33) to these results.

\renewcommand{\arraystretch}{1}

\begin{table}[ht]
\begin{center}
\begin{tabular}{|c|c|c|c|c|c|c|}
\hline & & & & & & \\
&  $({1\atop c})({d \atop 2})$ & $({1\atop c})({d \atop 3})$ & $({2\atop c})({d \atop 3})$ 
 & $({2\atop c})({d \atop 1})$ & $({3\atop c})({d \atop 1})$  & $({3\atop c})({d \atop 2})$ 
\\  & & & & & & \\ \hline & & & & & & \\ 
$({1\atop a})({b\atop 2})$ & $0$ & $0$ & $0$ & $\scriptstyle{\delta_a^b\delta_c^d y(T^2_c)}
$ & $0$ & $0$  \\  & & & & & & \\ \hline & & & & & & \\
$({1\atop a})({b\atop 3})$ & $0$ & $0$ & $0$ & $0$ & $\scriptstyle{\delta_a^b\delta_c^d 
y(T^3_c)}$ & $0$ \\  & & & & & & \\ \hline & & & & & & \\ 
$({2\atop a})({b\atop 3})$ & $0$ & $0$ & $0$ & $0$ & $0$ & 
$\scriptstyle{\delta_a^b\delta_c^d y(T^3_c)}$  \\ & & & & & & \\ \hline
 & & & & & & \\
$({2\atop a})({b\atop 1})$ & $\scriptstyle{\delta_a^b\delta_c^d y(T^2_c)}$ & $0$ & $0$ & 
$0$ & $0$ & $0$ \\  & & & & & & \\ \hline 
& & & & & & \\
$({3\atop a})({b\atop 1})$ & $0$ &  $\scriptstyle{\delta_a^b\delta_c^d y(T^3_c)}$ & $0$ & 
$0$ & $0$ &  $0$  \\  & & & & & & \\ \hline 
& & & & & & \\  
$({3\atop a})({b\atop 2})$ & $0$ & $0$ 
& $\scriptstyle{\delta_a^b\delta_c^d y(T^3_c)}$  & $0$ & $0$ &  $0$  \\
 & & & & & & \\ \hline 

\end{tabular}
\caption{The Riemann curvature $G^{({i\atop a})({b\atop j})({j\atop c})({d\atop i})}$ of 
the type $({\bf 5^*},{\bf 5},{\bf 5^*},{\bf 5})$ with $i\ne j$.}
\end{center}
\end{table}

\begin{table}[ht]
\begin{center}
\begin{tabular}{|c|c|c|c|}
\hline & & & \\
&  $({1\atop c})({d \atop 1})$ & $({2\atop c})({d \atop 2})$ & $({3\atop c})({d \atop 3})$    
\\   & & & \\ \hline  & & & \\
$({1\atop a})({b\atop 1})$  & $\scriptstyle{(\delta_a^b\delta_c^d + 
\delta_a^d\delta_c^b)y(T^1_c)}$ & $\scriptstyle{\delta_a^d\delta_c^b y(T^2_c)}$ &  
$\scriptstyle{\delta_a^d\delta_c^b y(T^3_c)}$ \\    & & & \\ \hline  & & & \\
$({2\atop a})({b\atop 2})$ & $\scriptstyle{\delta_a^d\delta_c^b y(T^2_c)}$  & 
$\scriptstyle{(\delta_a^b\delta_c^d + \delta_a^d\delta_c^b)y(T^2_c)}$ & 
$\scriptstyle{\delta_a^d\delta_c^b y(T^3_c)}$
 \\   & & & \\ \hline  & & & \\
$({3\atop a})({b\atop 3})$ & $\scriptstyle{\delta_a^d\delta_c^b y(T^3_c)}$  &  
$\scriptstyle{\delta_a^d\delta_c^b y(T^3_c)}$ & $\scriptstyle{(\delta_a^b\delta_c^d + 
\delta_a^d\delta_c^b)y(T^3_c)}$ \\    & & & \\ \hline 
\end{tabular}
\caption{The Riemann curvature $G^{({i\atop a})({b\atop i})({j\atop c})({d\atop j})}$ of 
the type $({\bf 5^*},{\bf 5},{\bf 5^*},{\bf 5})$ }
\end{center}
\end{table}


\begin{table}[b]
\begin{center}
\begin{tabular}{|c|c|c|c|}
\hline  & & & \\
 & $({ef\atop 1})({2\atop gh})$ & $({ef\atop 1})({3\atop gh})$
& $({ef\atop 2})({3\atop gh})$   \\  & & & \\ \hline  & & & \\
$({ab\atop 2})({1\atop cd})$ & ${\delta^{ab}_{cd}\delta^{ef}_{gh}y(E^{ef}_1)\atop - 
\delta^{abef}_{cdgh}{y(E^{ef}_1)y(E^{gh}_2)\over y(T^3_e)}}$  & $0$ & $0$  \\   & & & \\ 
\hline  & & & \\
$({ab\atop 3})({1\atop cd})$ & $0$ & ${\delta^{ab}_{cd}\delta^{ef}_{gh}y(E^{ef}_1)\atop - 
\delta^{abef}_{cdgh}{y(E^{ef}_1)y(E^{gh}_3)\over y(T^2_e)}}$ &  $0$  \\  & & & \\ \hline  & 
& & \\
$({ab\atop 3})({2\atop cd})$ &  $0$ & $0$ & 
${\delta^{ab}_{cd}\delta^{ef}_{gh}y(E^{ef}_2)\atop - 
\delta^{abef}_{cdgh}{y(E^{ef}_2)y(E^{gh}_3)\over y(T^1_e)}}$  \\  & & & \\ \hline

\end{tabular}
\caption{The Riemann curvature $G^{({ab\atop j})({i\atop cd})({ef\atop i})({j\atop gh})}$ 
of  the type $({\bf 10},{\bf 10^*},{\bf 10},{\bf 10^*})$ with $i< j$.}
\end{center}
\end{table}

\begin{table}[ht]
\begin{center}
\begin{tabular}{|c|c|c|c|}
\hline  & & & \\
 & $({ef\atop 2})({1\atop gh})$ & $({ef\atop 3})({1\atop gh})$
& $({ef\atop 3})({2\atop gh})$   \\ & & & \\ \hline & & & \\
$({ab\atop 1})({2\atop cd})$ & ${\delta^{ab}_{cd}\delta^{ef}_{gh}y(E^{ef}_1)\atop - 
\delta^{abef}_{cdgh}{y(E^{ef}_2)y(E^{gh}_1)\over y(T^3_e)}}$  & $0$ & $0$  \\ & & & \\ 
\hline & & & \\
$({ab\atop 1})({3\atop cd})$ & $0$ & ${\delta^{ab}_{cd}\delta^{ef}_{gh}y(E^{ef}_1)\atop - 
\delta^{abef}_{cdgh}{y(E^{ef}_3)y(E^{gh}_1)\over y(T^2_e)}}$ &  $0$  \\ & & & \\ \hline & & 
& \\
$({ab\atop 2})({3\atop cd})$ &  $0$ & $0$ & 
${\delta^{ab}_{cd}\delta^{ef}_{gh}y(E^{ef}_2)\atop - 
\delta^{abef}_{cdgh}{y(E^{ef}_3)y(E^{gh}_2)\over y(T^1_e)}}$  \\ & & & \\ \hline 

\end{tabular}
\caption{The Riemann curvature $G^{({ab\atop j})({i\atop cd})({ef\atop i})({j\atop gh})}$ 
of  the type $({\bf 10},{\bf 10^*},{\bf 10},{\bf 10^*})$ with $i> j$.}
\end{center}
\end{table}

\renewcommand{\arraystretch}{1}

\begin{table}[ht]
\begin{center}
\begin{tabular}{|c|c|c|c|}
\hline
& & & \\
 & $({ef\atop 1})({1\atop gh})$ & $({ef\atop 2})({2\atop gh})$
& $({ef\atop 3})({3\atop gh})$  \\
 & & &  \\ \hline  & & & \\
$({ab\atop 1})({1\atop cd})$ & 
$\scriptstyle {(\delta^{a\; b}_{c[h}\delta^{e\; f}_{g]d} 
+ \delta^{a\; b}_{d[g}\delta^{e\; f}_{h]c})y(E^{ef}_1)}$ 
& ${\delta^{ab}_{gh}\delta^{ef}_{cd}y(E^{ef}_1)\atop 
 - \delta^{abef}_{cdgh}{y(E^{ef}_2)y(E^{ab}_1)\over y(T^3_e)}}$
& ${\delta^{ab}_{gh}\delta^{ef}_{cd}y(E^{ef}_1)\atop 
 - \delta^{abef}_{cdgh}{y(E^{ef}_3)y(E^{ab}_1)\over y(T^2_e)}}$  \\
& & &  \\ \hline  & & & \\

$({ab\atop 2})({2\atop cd})$ 
& ${\delta^{ab}_{gh}\delta^{ef}_{cd}y(E^{ef}_1)\atop 
 - \delta^{abef}_{cdgh}{y(E^{ef}_1)y(E^{ab}_2)\over y(T^3_e)}}$
& $\scriptstyle {(\delta^{a\; b}_{c[h}\delta^{e\; f}_{g]d} 
+ \delta^{a\; b}_{d[g}\delta^{e\; f}_{h]c})y(E^{ef}_2)}$ 
& ${\delta^{ab}_{gh}\delta^{ef}_{cd}y(E^{ef}_2)\atop 
 - \delta^{abef}_{cdgh}{y(E^{ef}_3)y(E^{ab}_2)\over y(T^1_e)}}$  \\
 & & &  \\ \hline & & & \\  
$({ab\atop 3})({3\atop cd})$ 
& ${\delta^{ab}_{gh}\delta^{ef}_{cd}y(E^{ef}_1)\atop 
 - \delta^{abef}_{cdgh}{y(E^{ef}_1)y(E^{ab}_3)\over y(T^2_e)}}$
& ${\delta^{ab}_{gh}\delta^{ef}_{cd}y(E^{ef}_2)\atop 
 - \delta^{abef}_{cdgh}{y(E^{ef}_2)y(E^{ab}_3)\over y(T^1_e)}}$ 
& $\scriptstyle {(\delta^{a\; b}_{c[h}\delta^{e\; f}_{g]d} 
+ \delta^{a\; b}_{d[g}\delta^{e\; f}_{h]c})y(E^{ef}_3)}$  \\
& & & 
 \\ \hline
\end{tabular}
\caption{The Riemann curvature $G^{({ab\atop i})({i\atop cd})({ef\atop k})({k\atop gh})}$ 
of  the type $({\bf 10},{\bf 10^*},{\bf 10},{\bf 10^*})$.}
\end{center}
\end{table}


\renewcommand{\arraystretch}{1}

\begin{table}[ht]
\begin{center}
\begin{tabular}{|c|c|c|c|c|c|c|}
\hline & & & & & & \\
& $({1\atop 2})({3 \atop 1})$  & $({1\atop 3})({2\atop 1})$ & $({1\atop 3})({3\atop 2})$ 
 & $({2\atop 3})({3 \atop 1})$ & $({1\atop 2})({3\atop 2})$  & $({2\atop 3})({2 \atop 1})$   
\\  & & & & & & \\ \hline & & & & & & \\
$({1\atop 2})({3\atop 1})$ & $0$ & $\scriptstyle{y(T^1_2)}$ & $0$ & $0$ & $0$ & $0$  \\ 
& & & & & & \\ \hline & & & & & & \\
$({1\atop 3})({2\atop 1})$ & $\scriptstyle{y(T^1_2)}$ & $0$ & $0$ & $0$ & $0$ & $0$ \\ 
& & & & & & \\  \hline & & & & & & \\
$({1\atop 3})({3\atop 2})$ & $0$ & $0$ & $0$ & $\scriptstyle{y(T^2_3)}$ & $0$ & $0$  \\ 
 & & & & & & \\ \hline & & & & & & \\
$({2\atop 3})({3\atop 1})$ & $0$ & $0$ & $\scriptstyle{y(T^2_3)}$ & $0$ & $0$ & $0$ \\ 
 & & & & & & \\ \hline & & & & & & \\
$({1\atop 2})({3\atop 2})$ & $0$ &  $0$ & $0$ & $0$ & $0$ &  
$-{y(T^2_3)y(T^1_2)\over y(T^1_3)}$  \\  
& & & & & & \\ \hline & & & & & & \\
$({2\atop 3})({2\atop 1})$ & $0$ & $0$ & $0$  & $0$ & $-{y(T^2_3)y(T^1_2)\over y(T^1_3)}$ &  
$0$  \\ & & & & & & \\ \hline

\end{tabular}
\caption{The Riemann curvature $G^{({i\atop j})({k\atop l})({m\atop l})
({p\atop q})}$ of the type $({\bf 1},{\bf 1},{\bf 1},{\bf 1})$ with $y([T^i_j,T^k_l]) \ne 
0$.}
\end{center}
\end{table}

\begin{table}[ht]
\begin{center}
\begin{tabular}{|c|c|c|c|}
\hline & & & \\
 & $({1\atop 2})({2\atop 1})$ & \hspace{0.4cm}$({1\atop 3})({3\atop 1})$\hspace{0.4cm}
& $({2\atop 3})({3\atop 2})$   \\ & & & \\ \hline & & & \\
$({1\atop 2})({2\atop 1})$ 
& $\scriptstyle{2y(T^1_2)}$ & $\scriptstyle{y(T^1_2)}$ & $-{y(T^2_3)y(T^1_2)\over y(T^1_3)}
$ \\ 
& & & \\ \hline & & & \\
$({1\atop 3})({3\atop 1})$ & $\scriptstyle{y(T^1_2)}$ & $\scriptstyle{2y(T^1_3)}$ & 
$\scriptstyle{y(T^2_3)}$ \\ 
& & & \\ \hline & & & \\
$({2\atop 3})({3\atop 2})$ & $-{y(T^2_3)y(T^1_2)\over y(T^1_3)}$ & $\scriptstyle{y(T^2_3)}$
& $\scriptstyle{2y(T^2_3)}$  \\ & & & \\ \hline

\end{tabular}
\caption{The Riemann curvature $G^{({i\atop j})({j\atop i})({k\atop l})({l\atop k})}$ of 
the type $({\bf 1},{\bf 1},{\bf 1},{\bf 1})$.}
\end{center}
\end{table}


\renewcommand{\arraystretch}{1}

\begin{table}[ht]
\begin{center}
\begin{tabular}{|c|c|c|c|}
\hline & & & \\
 & $({cd\atop 1})({2\atop ef})$ & $({cd\atop 1})({3\atop ef})$
& $({cd\atop 2})({3\atop ef})$  \\
  & & &  \\ \hline  & & & \\
$({1\atop a})({b\atop 2})$ 
& ${-\delta^b_a\delta^{cd}_{ef}y(E^{cd}_1)\atop 
-\delta^{\; b}_{[e}\delta^{c\; d}_{f]a}
{y(E^{cd}_1)y(E^{ef}_2)\over y(E^a)}}$
& $0$ & $0$  \\ 
 & & & \\ \hline & & & \\
$({1\atop a})({b\atop 3})$ 
& $0$ 
& ${-\delta^b_a\delta^{cd}_{ef}y(E^{cd}_1)  \atop
 -\delta^{\; b}_{[e}\delta^{c\; d}_{f]a}
{y(E^{cd}_1)y(E^{ef}_3)\over y(E^a)}}$

& $0$  \\ 
 & & & \\ \hline & & & \\
$({2\atop a})({b\atop 3})$ 
& $0$ & $0$ 
& ${-\delta^b_a\delta^{cd}_{ef}y(E^{cd}_2)\atop 
-\delta^{\; b}_{[e}\delta^{c\; d}_{f]a}
{y(E^{cd}_2)y(E^{ef}_3)\over y(E^a)}}$  \\ 
 & & & \\ \hline 
\end{tabular}
\caption{The Riemann curvature $G^{({i\atop a})({b\atop j})({cd\atop i})({j\atop ef})}$ of  
the type $({\bf 5^*},{\bf 5},{\bf 10},{\bf 10^*})$ with $i< j$.}
\end{center}
\end{table}

\renewcommand{\arraystretch}{1}

\begin{table}[ht]
\begin{center}
\begin{tabular}{|c|c|c|c|}
\hline & & & \\
 & $({cd\atop 2})({1\atop ef})$ & $({cd\atop 3})({1\atop ef})$
& $({cd\atop 3})({2\atop ef})$  \\
  & & &  \\ \hline  & & & \\
$({2\atop a})({b\atop 1})$ 
& ${-\delta^b_a\delta^{cd}_{ef}y(E^{cd}_1)\atop 
-\delta^{\; b}_{[e}\delta^{c\; d}_{f]a}
{y(E^{cd}_2)y(E^{ef}_1)\over y(E^a)}}$
& $0$ & $0$  \\ 
 & & & \\ \hline & & & \\
$({3\atop a})({b\atop 1})$ 
& $0$ 
& ${-\delta^b_a\delta^{cd}_{ef}y(E^{cd}_1) \atop
 -\delta^{\; b}_{[e}\delta^{c\; d}_{f]a}
{y(E^{cd}_3)y(E^{ef}_1)\over y(E^a)}}$

& $0$  \\ 
 & & & \\ \hline & & & \\
$({3\atop a})({b\atop 2})$ 
& $0$ & $0$ 
& ${-\delta^b_a\delta^{cd}_{ef}y(E^{cd}_2)\atop 
-\delta^{\; b}_{[e}\delta^{c\; d}_{f]a}
{y(E^{cd}_3)y(E^{ef}_2)\over y(E^a)}}$  \\ 
 & & & \\ \hline 
\end{tabular}
\caption{The Riemann curvature $G^{({i\atop a})({b\atop j})({cd\atop i})({j\atop ef})}$ of  
the type $({\bf 5^*},{\bf 5},{\bf 10},{\bf 10^*})$ with $i> j$.}
\end{center}
\end{table}

\renewcommand{\arraystretch}{1}

\begin{table}[ht]
\begin{center}
\begin{tabular}{|c|c|c|c|}
\hline & & & \\
& $({cd\atop 1})({1\atop ef})$
& $({cd\atop 2})({2\atop ef})$
& $({cd\atop 3})({3\atop ef})$  \\
  & & &  \\ \hline  & & & \\
$({1\atop a})({b\atop 1})$ 
& $\scriptstyle{-\delta^{\; b}_{[e}\delta^{c\; d}_{f]a}
{y(E^{cd}_1)y(T^1_a)\over y(E^a)}}$
& $\scriptstyle{\delta^{bcd}_{aef}y(E^{cd}_2)}$ 
& $\scriptstyle{\delta^{bcd}_{aef}y(E^{cd}_3)}$  \\
& & & \\ \hline & & & \\
$({2\atop a})({b\atop 2})$ 
& $\scriptstyle{\delta^{bcd}_{aef}y(E^{cd}_1)}$ 
& $\scriptstyle{-\delta^{\; b}_{[e}\delta^{c\; d}_{f]a}
{y(E^{cd}_2)y(T^2_a)\over y(E^a)}}$
& $\scriptstyle{\delta^{bcd}_{aef}y(E^{cd}_3)}$  \\
& & & \\ \hline & & & \\
$({3\atop a})({b\atop 3})$ 
& $\scriptstyle{\delta^{bcd}_{aef}y(E^{cd}_1)}$ 
& $\scriptstyle{\delta^{bcd}_{aef}y(E^{cd}_2)}$
& $\scriptstyle{-\delta^{\; b}_{[e}\delta^{c\; d}_{f]a}
{y(E^{cd}_3)y(T^3_a)\over y(E^a)}}$    \\
& & &  \\ \hline 

\end{tabular}
\caption{The Riemann curvature $G^{({i\atop a})({b\atop i})({cd\atop k})({k\atop ef})}$ of  
the type $({\bf 5^*},{\bf 5},{\bf 10},{\bf 10^*})$.}
\end{center}
\end{table}


\renewcommand{\arraystretch}{1}

\begin{table}[ht]
\begin{center}
\begin{tabular}{|c|c|c|c|c|c|c|}
\hline  & & & & & & \\
& $({1\atop a})({b \atop 2})$ & $({1\atop a})({b \atop 3})$ & $({2\atop a})({b \atop 3})$ & 
$({2\atop a})({b \atop 1})$ & $({3\atop a})({b \atop 1})$  & $({3\atop a})({b\atop 2})$ \\  
& & & & & & \\ \hline  & & & & & & \\
$({1\atop 2})({3 \atop 1})$  &  $0$ & $0$ 
& $-\delta^b_a {y(T^3_a)y(T^1_2)\over y(T^1_a)}$  
& $0$ & $0$ & $0$   \\   & & & & & & \\ \hline  & & & & & & \\
$({1\atop 3})({2\atop 1})$ & $0$ & $0$ & $0$  & $0$ & $0$ 
&$-{y(T^3_a)y(T^1_2)\over y(T^1_a)}$   \\   & & & & & & \\ \hline  & & & & & & \\
$({1\atop 3})({3\atop 2})$ & $0$ & $0$ & $0$ & $\scriptstyle{\delta^b_a y(T^2_3)}$ & $0$ & 
$0$ \\ & & & & & & \\ \hline   & & & & & & \\
$({2\atop 3})({3\atop 1})$ & $\scriptstyle{\delta^b_a y(T^2_3)}$ & $0$ & $0$ & $0$ & $0$ & 
$0$  \\   & & & & & & \\ \hline  & & & & & & \\
$({1\atop 2})({3\atop 2})$ & $0$ &  $0$ & $0$ & $0$ & $0$ &  
$0$  \\   & & & & & & \\ \hline  & & & & & & \\
$({2\atop 3})({2\atop 1})$ & $0$ & $0$ & $0$  & $0$ & $0$ &  $0$  \\  & & & & & & \\ \hline  

\end{tabular}
\caption{The Riemann curvature $G^{({i\atop j})({k\atop l})({m\atop a})
({b\atop n})}$ of the type $({\bf 1},{\bf 1},{\bf 5^*},{\bf 5})$ with $y([T^i_j,T^k_l]) \ne 
0$.}
\end{center}
\end{table}

\begin{table}[ht]
\begin{center}
\begin{tabular}{|c|c|c|c|}
\hline  & & & \\
 & $({1\atop a})({b\atop 1})$ & $({2\atop a})({b\atop 2})$
& $({3\atop a})({b\atop 3})$   \\ & & &  \\ \hline &  & & \\
$({1\atop 2})({2\atop 1})$ 
& $\scriptstyle{\delta^b_a y(T^1_2)}$
& $-\scriptstyle{\delta^b_a }{y(T^1_2)(y(T^2_a)\over y(T^1_a)}$ 
& $0$  \\  & & &  \\ \hline & & & \\
$({1\atop 3})({3\atop 1})$ 
& $\scriptstyle{\delta^b_a y(T^1_3)}$
& $0$ 
& $-\scriptstyle{\delta^b_a }{y(T^1_3)y(T^3_a)\over y(T^1_a)}$ 
  \\ &  & & \\ \hline &  & & \\
$({2\atop 3})({3\atop 2})$ 
& $0$ 
& $\scriptstyle{\delta^b_a y(T^2_3)}$
& $-\scriptstyle{\delta^b_a }{y(T^2_3)y(T^3_a)\over y(T^2_a)}$ 
 \\ & & &  \\ \hline

\end{tabular}
\caption{The Riemann curvature $G^{({i\atop j})({j\atop i})({k\atop a})
({b\atop k})}$ of the type $({\bf 1},{\bf 1},{\bf 5^*},{\bf 5})$.}
\end{center}
\end{table}


\renewcommand{\arraystretch}{1}

\begin{table}[ht]
\begin{center}
\begin{tabular}{|c|c|c|c|c|c|c|}
\hline & & & & & & \\
 & $({ab\atop 2})({1\atop cd})$ & $({ab\atop 3})({1\atop cd})$
 & $({ab\atop 3})({2\atop cd})$ & $({ab\atop 1})({2\atop cd})$
 & $({ab\atop 1})({3\atop cd})$ & $({ab\atop 2})({3\atop cd})$ \\ 
 & & & & & & \\
 \hline  & & & & & & \\
 $({1\atop 2})({3\atop 1})$ & $0$ & $0$ 
 & $\scriptstyle{\delta^{ab}_{cd}y(T^1_2)}$ & $0$ & $0$ & $0$  \\
 & & & & & & \\
 \hline & & & & & & \\
 $({1\atop 3})({2\atop 1})$ & $0$ & $0$  & $0$ & $0$ & $0$
 & $\scriptstyle{\delta^{ab}_{cd}y(T^1_2)}$   \\
 & & & & & & \\
 \hline  & & & & & & \\
 $({1\atop 3})({3\atop 2})$ & $0$ & $0$ & $0$ 
 & $\scriptstyle{-\delta^{ab}_{cd}}{y(E^{ab}_1)y(T^2_3) \over y(E^{ab}_3)}$ & $0$ & $0$  \\
 & & & & & & \\
  \hline  & & & & & & \\
 $({2\atop 3})({3\atop 1})$ 
 & $\scriptstyle{-\delta^{ab}_{cd}}{y(E^{ab}_1)y(T^2_3) \over y(E^{ab}_3)}$
 & $0$ & $0$ & $0$  & $0$ & $0$  \\
 & & & & & & \\
 \hline & & & & & & \\
$({1\atop 2})({3\atop 2})$ & $0$ & $0$ 
 & $0$ & $0$ & $0$ & $0$  \\
 & & & & & & \\
 \hline & & & & & & \\
$({2\atop 3})({2\atop 1})$ & $0$ & $0$ 
 & $0$ & $0$ & $0$ & $0$  \\
 & & & & & & \\
 \hline

\end{tabular}
\caption{The Riemann curvature $G^{({i\atop j})({k\atop l})({ab \atop n})
({m\atop cd})}$ of the type $({\bf 1},{\bf 1},{\bf 10},{\bf 10^*})$ with $y([T^i_j,T^k_l]) 
\ne 0$.}
\end{center}
\end{table}

\clearpage

\renewcommand{\arraystretch}{1}

\begin{table}[ht]
\begin{center}
\begin{tabular}{|c|c|c|c|}
\hline  & & & \\
 & $({ab\atop 1})({1\atop cd})$ & $({ab\atop 2})({2\atop cd})$
& $({ab\atop 3})({3\atop cd})$   \\ & & &  \\ \hline &  & & \\
$({1\atop 2})({2\atop 1})$ 
& $\scriptstyle{-\delta^{ab}_{cd}}
 {y(T^1_2)y(E^{ab}_1)\over y(E^{ab}_2)} $
& $\scriptstyle{\delta^{ab}_{cd}y(T^1_2)}$
& $0$  \\ 
  & & &  \\ \hline & & & \\
$({1\atop 3})({3\atop 1})$ 
& $\scriptstyle{-\delta^{ab}_{cd}}
 {y(T^1_3)y(E^{ab}_1)\over y(E^{ab}_3)} $
& $0$ 
& $\scriptstyle{\delta^{ab}_{cd}y(T^1_3)}$  \\ 
& & &  \\ \hline & & & \\
$({2\atop 3})({3\atop 2})$ 
& $0$ 
& $\scriptstyle{-\delta^{ab}_{cd}}
 {y(T^2_3)y(E^{ab}_2)\over y(E^{ab}_3)} $
& $\scriptstyle{\delta^{ab}_{cd}y(T^2_3)}$
 \\ 
  & & &  \\ \hline

\end{tabular}
\caption{The Riemann curvature $G^{({i\atop j})({j\atop i})({ab \atop i})
({j\atop cd})}$ of the type $({\bf 1},{\bf 1},{\bf 10},{\bf 10^*})$.}
\end{center}
\end{table}

\section{Conclusions}
\setcounter{equation}{0}

In this paper we have discussed the reducible K\"ahler coset space $G/S\otimes \{U(1)\}^k$ 
in the geometrical approach generalizing the arguments in ref.\cite{9}. We have expressed 
the Riemann curvature of the coset space in terms of the Killing vectors  as (4.32). It is 
the most important formula in this paper. 
We have been then interested in the four-fermi coupling of the  supersymmetric non-linear 
$\sigma$-model on  $G/S\otimes \{U(1)\}^k$, to the leading order of ${1\over f}$. It is 
given by evaluating the Riemann curvature at the  origin of the coset space. We have 
 established the group theoretical method to do this by using the formula (4.32). Otherwise 
the calculation  would be too complicated. Concrete calculations have been done for 
$SU(3)/\{U(1)\}^2$ and $E_7/SU(5)\otimes \{U(1)\}^3$. The results of  the last K\"ahler 
coset space is phenomelogically interesting, since they give  four-fermi coupling constants 
among the three families of ${\bf 10} + {\bf 5^*} + {\bf 1}$ of $SU(5)$  in the  
supersymmetric non-linear $\sigma$-model on $E_7/SU(5)\otimes \{U(1)\}^3$. Among them those 
involving the three families of  right-handed neutrinos  are particularly interesting and 
have been given in Tables 8,9,13$\sim$16.  The dependence of the three $U(1)$-charges of 
the NG pseudo fermions are explicit in these results. Of course, we may take another set of 
$U(1)$s, say $Q^1, Q^2, Q^3$, than $T,Q,Q'$, for instance, 
 those which remain unbroken in the breaking process
$$
E_7 \stackrel{U(1)}\longrightarrow{E_6}
 \stackrel{U(1)}\longrightarrow{SO(10)}
  \stackrel{U(1)}\longrightarrow{SU(5)}
$$
 as in ref.\cite{5}. The results given by Tables 3$\sim$16 are still valid if one defines 
the the Y-charge as
 $$
 Y = \alpha Q^1 + \beta Q^2 + \gamma Q^3.
 $$
 and replaces Table 2 for  $y(X^{\ov i})$ by a new table with $Q^1, Q^2, Q^3$. It is 
desired to carry out a phenomelogical study by tuning the three arbitrary parameters
$\alpha, \beta, \gamma$.

\vspace{2cm}
\noindent
{\Large\bf Acknowledgements}

The author would like to thank T. Yanagida for reviving the interest in the subject. He is 
grateful to T. Kugo for the valuable discussions on the reducible K\"ahler coset space and 
reading the manuscript. 

\newpage

\appendix

\apsect{A}

\setcounter{equation}{0}

We derive the useful formulae for the calculation in Subsection {\bf 4.2}. We start with 
covariantization of the Lie-algebra (2.14):
$$
 R^{A\alpha} R^{B\beta}_{\ \  ;\alpha} - R^{B\alpha} R^{A\beta}_{\ \ ;\alpha}
 \ \ = \ \ f^{ABC}R^{C\beta}.  
$$
Multiplying both sides by $(R_\gamma\Delta)^B$ or $(R_{\ov\gamma}\Delta)^B$
and lower the index $\beta$ to get 
\begin{eqnarray}
R^{A\alpha}\cdot R_\gamma\Delta R_{\ov\beta,\alpha} - R^A_{\ov\beta,\gamma}
 & =&  f^{ABC}(R_\gamma\Delta)^B R^C_{\ov\beta},  
\end{eqnarray}
or
\begin{eqnarray}
R^{A\alpha}\cdot R_{\ov\gamma}\Delta R_{\ov\beta,\alpha} &=& 
f^{ABC}(R_{\ov\gamma}\Delta)^B R^C_{\ov\beta},
\end{eqnarray}
by using (4.23). Noting that 
\begin{eqnarray}
R_\gamma\Delta R_{\ov\beta,\alpha} &=& -R_\gamma\Delta_{,\alpha} R_{\ov\beta},
\end{eqnarray}
and
\begin{eqnarray}
R_{\ov\gamma}\Delta R_{\ov\beta,\alpha} \ \  = \ \
- R_{\ov\gamma}\Delta R_{\alpha,\ov\beta}\ \ = \ \
R_{\ov\gamma}\Delta_{,\ov\beta} R_{\alpha},
\end{eqnarray}
by (4.24) and (2.15), we write (A.1) and (A.2) respectively as
\begin{eqnarray}
R^A_{\ov\beta,\gamma} &=& - R^{A\alpha}\cdot R_\gamma\Delta_{,\alpha}R_{\ \ov\beta} 
 - f^{ABC}(R_\gamma \Delta)^B R^C_{\ \ov\beta}  
\end{eqnarray}
and
\begin{eqnarray}
R^{A\alpha}\cdot R_{\ov\gamma}\Delta_{,\ov\beta} R_{\alpha} &=& 
f^{ABC}(R_{\ov\gamma}\Delta)^B R^C_{\ \ov\beta}.
\end{eqnarray}
Taking the complex conjugation of them gives 
\begin{eqnarray}
R^A_{\gamma,\ov\beta} &=& - R^{A\ov\alpha}\cdot R_{\ov\beta}\Delta_{,\ov\alpha}
R_\gamma  -  f^{ABC}(R_{\ov\beta}\Delta)^B R^C_{\ \gamma}  
\end{eqnarray}
and
\begin{eqnarray}
R^{A\ov\alpha}\cdot R_{\gamma}\Delta_{,\beta} R_{\ov\alpha} &=& 
f^{ABC}(R_{\gamma}\Delta)^B R^C_{\ \beta}.
\end{eqnarray}
That (A.5) and (A.7) satisfy the Killing condition (2.15) can be easily checked
 by (4.17), i.e., 
$$
R^{A\alpha}\Delta_{,\alpha} + R^{A\ov\alpha}\Delta_{,\ov\alpha}
\ \ =\ \ i[T^A,\ \Delta ],
$$
with $(T^A)^{BC} = -i f^{ABC}$.
Multiplying both sides of (A.6) and (A.8) respectively by 
$(R_\eta \Delta)^A$ and $(R_{\ov\eta} \Delta)^A$ we get 
\begin{eqnarray}
R_{\ov\gamma}\Delta_{,\ov\beta}R_\eta  &=& 
 f^{ABC}(R_{\ov\gamma}\Delta)^A R^B_{\ \ov\beta}(R_\eta\Delta)^C
\end{eqnarray}
and
\begin{eqnarray}
R_{\gamma}\Delta_{,\beta}R_{\ov\eta} &=&
 f^{ABC}(R_{\gamma}\Delta)^A R^B_{\ \beta}(R_{\ov\eta}\Delta)^C
\end{eqnarray}
owing to (4.23).


\apsect{B}

\setcounter{equation}{0}

We will check the symmetry property 
$R_{\alpha\ov\sigma\beta\ov\delta}=  R_{\alpha\ov\delta\beta\ov\sigma}$
 of the r.h.s. of (4.32). Put it in the form 
\begin{eqnarray}
 R_{\alpha\ov\sigma\beta\ov\delta} 
  & = & [ f^{ABC}(R_\alpha \Delta)^A(R_{\ov\delta}\Delta)^B R^{C\ov\eta}\cdot
      f^{DEF}(R_{\ov\sigma}\Delta)^D (R_\beta \Delta)^E R_{\ \ov\eta}   
     \nonumber  \\
 &+ & f^{ABC}(R_\alpha \Delta)^A(R_{\ov\sigma}\Delta)^B R^{C\ov\eta}\cdot
      f^{DEF}(R_{\ov\delta} \Delta)^D (R_\beta \Delta)^E R_{\ \ov\eta}^F ]
  \nonumber\\
&+ & [  f^{ABC}(R_\alpha \Delta)^A(R_{\ov\sigma}\Delta)^B \cdot
      f^{CDE}R^D_{\ \beta}( R_{\ov\delta}\Delta)^E
      \nonumber \\
& - &  R_\alpha \Delta_{,\beta} R_{\ov\eta}\cdot 
   R_{\ov\sigma} \Delta_{,\ov\delta} R^{\ov \eta} ], 
\end{eqnarray}
remembering (4.31). The first bracket is already symmetric under the interchange of 
$\ov\sigma$ and $\ov\delta$. Therefore we are left with the second bracket to exsamine. The 
anti-symmetric sum of its first piece by interchanging $\ov\sigma$ and $\ov\delta$ becomes 
\begin{eqnarray}
  & &f^{ABC}(R_\alpha \Delta)^A(R_{\ov\sigma}\Delta)^B \cdot
      f^{CDE}R^D_{\ \beta}( R_{\ov\delta}\Delta)^E  \nonumber\\
  & & \hspace{1cm}- f^{ABC}(R_\alpha \Delta)^A(R_{\ov\delta}\Delta)^B \cdot
      f^{CDE}R^D_{\ \beta}( R_{\ov\sigma}\Delta)^E   \nonumber\\
  & &   = f^{ABC}(R_\alpha\Delta)^A R^B_{\ \beta}\cdot 
f^{CDE}(R_{\ov\sigma}\Delta)^D(R_{\ov\delta}\Delta)^E,
\end{eqnarray}
by using the Jacobi identity of the structure constants (4.15). On the other hand that of 
the second piece is given by 
\begin{eqnarray}
  & & (R_{\ov\sigma}\Delta_{,\ov\delta}R_\eta - R_{\ov\delta}\Delta_{,\ov\sigma}R_\eta)
\cdot R_\alpha\Delta_{,\beta}R^\eta   \nonumber\\
& & \hspace{2cm} = f^{ABC}(R_{\ov\sigma}\Delta)^A(R_{\ov\delta}\Delta)^B \cdot 
f^{CDE}(R_\alpha\Delta)^D R^E_{\ \beta}.
\end{eqnarray}
This is easily checked as follows. Note at first that 
$$
R_{\ov\sigma}\Delta_{,\ov\delta}R_\eta + R_{\ov\sigma}\Delta R_{\eta,\ov\delta} \ \ = \ \ 
0,
$$
from (4.24). Then plug (A.7) in this to find 
$$
R_{\ov\sigma}\Delta_{,\ov\delta}R_\eta - R_{\ov\delta}\Delta_{,\ov\sigma}R_\eta
 \ \ =\ \ f^{ABC}(R_{\ov\sigma}\Delta)^A(R_{\ov\delta}\Delta)^B R^C_{\ \eta}.
$$
Multiplying both sides by $R_\alpha \Delta_{,\beta}R^\eta$ and using (A.8) yields (B.3).
From (B.2) and (B.3) the second bracket of (B.1) is also symmetric under the interchange of 
$\ov\sigma$ and $\ov\delta$. Thus we have $R_{\alpha\ov\sigma\beta\ov\delta}=  
R_{\alpha\ov\delta\beta\ov\sigma}$.

Next we examine the symmetry property $R_{\alpha\ov\sigma\beta\ov\delta}=  
R_{\beta\ov\sigma\alpha\ov\delta}$. We rewrite the Riemann curvature (4.29) as
\begin{eqnarray}
 R_{\alpha\ov\sigma\beta\ov\delta} & = & 
 -(R_\alpha \Delta)_{,\ov\delta} R^{\ov\eta}
     \cdot R_{\ov\sigma} \Delta_{,\ov\eta} R_{\beta} \nonumber\\
  &+ &   f^{ABC}(R_\alpha \Delta)^A
       R^B_{\  \ov\delta,\beta}
       (R_{\ov\sigma}\Delta)^C   \\
   & -&  f^{ABC}(R_\alpha \Delta)^A
       R^B_{\  \beta}
       (R_{\ov\sigma}\Delta_{,\ov\delta})^C ,  \nonumber
\end{eqnarray}
by using the Killing condition (2.15). Then with (A.5),(A.9),(A.10),(4.30) and (4.31) it 
becomes 
\begin{eqnarray}
R_{\alpha\ov\sigma\beta\ov\delta} 
  &=& [ f^{ABC}(R_\alpha \Delta)^A(R_{\ov\delta}\Delta)^B R^{C\ov\eta}\cdot
      f^{DEF}(R_{\ov\sigma}\Delta)^D (R_\beta \Delta)^E R_{\ \ov\eta}^F
        \nonumber \\
 &+ & f^{ABC}(R_\beta \Delta)^A(R_{\ov\delta} \Delta)^B  R^{C \ov\eta}\cdot 
      f^{DEF}(R_{\ov\sigma}\Delta)^D(R_\alpha \Delta)^E R_{\ \ov\eta}^F ]
    \nonumber \\
 &+ & [ f^{ABC}(R_\alpha \Delta)^A(R_{\ov\sigma}\Delta)^B \cdot
      f^{CDE}(R^D_{\ \beta}\Delta)R^E_{\ \ov\delta}
      \nonumber\\
 & -&  f^{ABC}(R_\alpha \Delta)^A R^B_{\ \beta} (R_{\ov \eta}\Delta)^C \cdot
   f^{DEF}(R_{\ov\sigma}\Delta)^D R^E_{\ \ov\delta} (R^{\ov\eta}\Delta)^F  ]. 
\end{eqnarray}
The first bracket is symmetric under the interchange of $\alpha$ and $\beta$. The symmetry 
of the second bracket can be shown similarly to the previous demonstraration of 
$R_{\alpha\ov\sigma\beta\ov\delta}=  R_{\alpha\ov\delta\beta\ov\sigma}$.

\apsect{C}

\setcounter{equation}{0}

We show how to evaluate the Riemann curvature $G^{({ab \atop j})({i \atop cd})({ef \atop 
l}) ({k \atop gh})}$ of the type $({\bf 10},{\bf 10}^*,{\bf 10},{\bf 10}^*)$ by (5.3) and 
(5.13).
 For
$i>j$ we have $y([E^{ab}_j,E^i_{cd}]) < 0$. By (5.8) non-trivial components of the Riemann 
curvature are
\begin{eqnarray}
G^{({ab \atop j})({i \atop cd})({ef \atop i})
({j \atop gh})} & = & tr([E^{ab}_j,E^i_{cd}][E^{ef}_i,E^j_{gh}])
 ( y(E^{ef}_i)- y(T^j_i) )   \nonumber \\
&-& tr([E^{ab}_j,E^{ef}_i][E^i_{cd},E^j_{gh}])
 {y(E^{ef}_i)y(E^{gh}_j) \over y([E^{ab}_j,E^{ef}_i])} \nonumber \\
&=& \delta^{ab}_{cd}\delta^{ef}_{gh}y(E^{ef}_j) 
-\delta^{abef}_{cdgh}{y(E^{ef}_i)y(E^{gh}_j) \over y([E^{ab}_j,E^{ef}_i])}.  \nonumber
\end{eqnarray}
For $i<j$ we have $y([E^{ab}_j,E^i_{cd}]) > 0$. By (5.7) non-trivial components of the 
Riemann curvature are 
\begin{eqnarray}
G^{({ab \atop j})({i \atop cd})({ef \atop i})
({j \atop gh})} & = & tr([E^{ab}_j,E^i_{cd}][E^{ef}_i,E^j_{gh}])
 y(E^{ef}_i)  \nonumber\\
 &-& tr([E^{ab}_j,E^{ef}_i][E^i_{cd},E^j_{gh}])
 {y(E^{ef}_i)y(E^{gh}_j) \over y([E^{ab}_j,E^{ef}_i])} \nonumber\\
&=& \delta^{ab}_{cd}\delta^{ef}_{gh}y(E^{ef}_i) 
-\delta^{abef}_{cdgh}{y(E^{ef}_i)y(E^{gh}_j) \over y([E^{ab}_j,E^{ef}_i])}.  \nonumber
\end{eqnarray}
The same result is also obtained by interchanging the indices as 
 $$G^{({ab \atop j})({i \atop cd})({ef \atop i})
({j \atop gh})} = G^{({ef \atop i})({j \atop gh}){ab \atop j})({i \atop cd})}$$ and 
calculating it by (5.8). For $i=j$ we have $y([E^{ab}_j,E^i_{cd}]) = 0$. Non-trivial 
components of the Riemann curvature $G^{({ab \atop i})({i \atop cd})({ef \atop k})({k \atop 
gh})}$  are evaluated by (5.9).
For $k<i$ 
\begin{eqnarray}
G^{({ab \atop i})({i \atop cd})({ef \atop k})({k \atop gh})}
 &=& tr([E^{ab}_i,E^i_{cd}][E^{ef}_{k},E^{k}_{gh}])
  y(E^{ef}_{k})   \nonumber \\
&-& tr([E^{ab}_i,E^{ef}_{k}][E^i_{cd},E^{k}_{gh}])
{y(E^{ef}_{k})y(E^{gh}_{k}) \over y([E^{ab}_i,E^{ef}_{k}])}.\nonumber
\end{eqnarray}
By the formula 
$$
tr([E^{ab}_i,E^i_{cd}][E^{ef}_{k},E^{k}_{gh}])  = -\delta^{abef}_{cdgh}
 +  \delta^{ab}_{gh}\delta^{ef}_{cd}\ ,
$$
it becomes 
\begin{eqnarray}
G^{({ab \atop i})({i \atop cd})({ef \atop k})({k \atop ef})}
 &=& \delta^{ab}_{gh}\delta^{ef}_{cd}y(E^{ef}_k) - 
 \delta^{abef}_{cdgh}{y(E^{ab}_{i})y(E^{ef}_{k}) \over y([E^{ab}_i,E^{ef}_{k}])}. \nonumber
\end{eqnarray}
For $k>i$ 
\begin{eqnarray}
G^{({ab \atop i})({i \atop cd})({ef \atop k})({k \atop gh})}
 &=& \delta^{ab}_{gh}\delta^{ef}_{cd}y(E^{ef}_{i}) - 
 \delta^{abef}_{cdgh}{y(E^{ab}_{i})y(E^{ef}_{k}) \over y([E^{ab}_i,E^{ef}_{k}])}, \nonumber
\end{eqnarray}
by applying the symmetry property $$G^{({ab \atop i})({i \atop cd})({ef \atop k})({k \atop 
gh})} = G^{({ef \atop k})({k \atop gh})({ab \atop i})({i \atop cd})}$$ to the above result. 
Of course this can be obtained by a direct calculation. Finally for $k=i$ 
\begin{eqnarray}
G^{({ab \atop i})({i \atop cd})({ef \atop i})({i \atop gh})}
&=& tr([E^{ab}_i,E^i_{cd}][E^{ef}_i,E^i_{gh}])y(E^{ef}_i) \nonumber\\
&=& (\delta^{ab}_{ch}\delta^{ef}_{gd} + \delta^{ab}_{cg}\delta^{ef}_{dh}+
\delta^{ab}_{dh}\delta^{ef}_{cg}+\delta^{ab}_{gd}\delta^{ef}_{ch})y(E^{ef}_i). \nonumber
\end{eqnarray}

Other types of  the Riemann curvature are obtained similarly.

\end{document}